\documentclass[amsmath,amssymb,aps,nofootinbib,reprint,usenames,dvipsnames,floatfix]{revtex4-1}
\pdfoutput=1 
\usepackage[english]{babel}
\usepackage[utf8]{inputenc}
\usepackage[T1]{fontenc}

\usepackage{amsmath,amsfonts,amsthm,bm,bbm}
\usepackage{graphicx}
\usepackage{amssymb}
\usepackage{braket}
\usepackage{mathtools}
\usepackage[export]{adjustbox}
\usepackage{xcolor}
\usepackage{chngcntr}
\usepackage[inline]{enumitem}
\usepackage[normalem]{ulem}
\usepackage{float}
\usepackage[caption=false]{subfig}

\usepackage{tikz}
\usetikzlibrary{calc}

\usepackage{mathrsfs}

\usepackage{hyperref}

\newcommand\PSF{\mathrm{PSF}}
\DeclareMathOperator{\sinc}{sinc}
\newcommand\op[1]{\hat{#1}}

\newcommand\Trace[1]{\operatorname{Tr}\left(#1\right)}
\newcommand{\proj}[1]{\mbox{$|#1\rangle \!\langle #1 |$}}
\DeclareMathOperator{\diag}{diag}
\DeclareMathOperator{\rank}{rank}
\newcommand\Vect[1]{\vert #1 ) }
\newcommand\Vectr[1]{ ( #1  \vert }
\newcommand\vecb[1]{\mathrm{vecb}\left(#1\right)}

\newcommand\TracySingh{\odot}

\makeatletter
\renewcommand*\env@matrix[1][*\c@MaxMatrixCols c]{%
  \hskip -\arraycolsep
  \let\@ifnextchar\new@ifnextchar
  \array{#1}}
\makeatother

\begin{document}

\title{Quantum limits of localisation microscopy}

\author{Evangelia Bisketzi}
\author{Dominic Branford}
\author{Animesh Datta}

\affiliation{Department of Physics, University of Warwick, Coventry, CV4 7AL, United Kingdom}

\begin{abstract}
Localisation microscopy of multiple weak, incoherent point sources with possibly different intensities in one spatial dimension is equivalent to estimating the amplitudes of a classical mixture of coherent states of a simple harmonic oscillator.
This enables us to bound the multi-parameter covariance matrix for an unbiased estimator for the locations in terms of the quantum Fisher information matrix, which we obtained analytically.
In the regime of arbitrarily small separations we find it to be no more than rank two -- implying that no more than two independent parameters can be estimated irrespective of the number of point sources.
We use the eigenvalues of the classical and quantum Fisher information matrices to compare the performance of spatial-mode demultiplexing and direct imaging in localisation microscopy with respect to the quantum limits. 
\end{abstract}

\maketitle

\section{Introduction}

Precisely locating multiple single emitters is a key challenge in fluorescence microscopy. 
The process of estimating these locations depends on the quality of the image obtained by the microscope.
One of the major limitations to the image quality, known since Abbe and Rayleigh, lies in spatially resolving objects substantially smaller than half the wavelength of the light involved~\cite{BoWo99}.
Known as the Rayleigh limit or diffraction limit, it is a consequence of the diffraction of light due to its wave nature. 

Over the last couple of decades, ways to circumvent the Rayleigh limit in far-field fluorescence microscopy have been invented~\cite{comparison_microscopy}.
Confocal methods such as STED, RESOLFT, and SSIM~\cite{Hell:94,Heintzmann02,Gustafsson13081,Hofmann17565} use patterned illumination to spatially modulate the fluorescence pattern of emitters within a diffraction-limited region such that not all of them emit simultaneously, thereby achieving sub-Rayleigh resolution.
Other far-field methods such as PALM, fPALM and STORM~\cite{STORM,PALM,fluroscence_PALM} temporally modulate the fluorescence pattern of emitters with weak laser pulses stochastically such that only a low density of emitters are active within the Rayleigh limit at one time.
Repeating the process many times, images with sub-Rayleigh resolution are reconstructed from the measured positions of individual emitters.
These techniques, with resolution of tens of nanometers, have provided insights into biological processes at the cellular scale that were hitherto unattainable~\cite{diffraction_barrier}.

Though immensely powerful and impressive, none of these methods seek to extract all the information available in the emitted light field.
As in conventional fluorescence microscopy these techniques use `direct imaging'---intensity measurements on the image plane---to extract information from the incident light.
That there is indeed more information in the light field to be extracted was shown by~\citet{TsaNL16}. Using methods from classical and quantum estimation theory, it was shown theoretically that two arbitrarily close incoherent point sources may be resolved, and that this may be achieved in practice using a spatial-mode demultiplexing (SPADE) measurement.
In the few years since, theoretical studies have considered different source arrangements or parameters of interest~\cite{Nair_superr_thermal,LuPi16,KeGu17,ChRaMaBa17,ReHr17,DuKeGu18} in one as well as in two and three spatial dimensions~\cite{TsaNai2D17,YuPra18,NaAd18,BaShWa18}.
Other theoretical studies have explored various detection systems that could achieve the ultimate precision in imaging or get close to it~\cite{nair_interferometric_2016,hom_heter_17,optimal_measurements_17,optimal_measurements_18}.
Several experiments have demonstrated some of the principles underlying these detection systems~\cite{tang_fault-tolerant_2016,achieving_optimal_res16,hom_het_experimental16,donohue_quantum-limited_2018,PSF_shaping18,parniak_beating_2018,mode_shorter_experim19,phase_sensitive_measur19}.
Advances in this area have been recently reviewed by \citet{tsang_resolving_2019}.

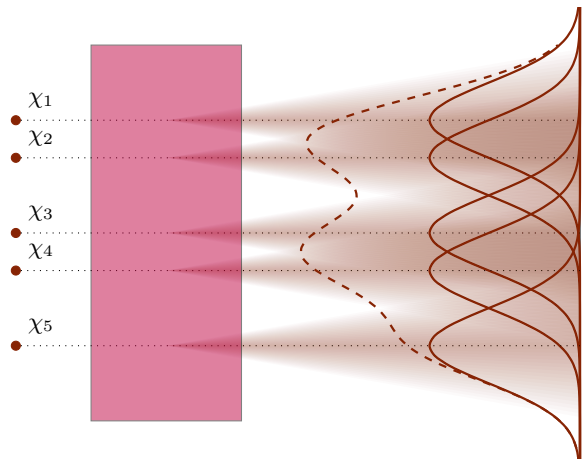
\begin{figure}[t]
\centering
\definecolor{myred}{rgb}{0.53,0.14,0.014}
\begin{tikzpicture}[pointsource/.style={circle,inner sep=0pt,minimum size=1.2mm},]
	\clip (-0.5,-0.5) rectangle (9.5,5.5);
	\coordinate (imageplane) at (7.5,0);
	\draw ($(imageplane)+(0,5)$) -- ($(imageplane)-(0,5)$);
	\foreach \ypos [count=\j] in {4,3.5,2.5,2,1}{
		\node (x-\j) [draw=myred,fill=myred,pointsource,label=above right:{$\chi_{\j}$}] at (0,\ypos) {};
		\draw [dotted] (x-\j) -- (x-\j -| imageplane);
		\draw [draw=myred,domain=-5:5,variable=\y,smooth,shift={(x-\j -| imageplane)},samples=75,line width=0.3mm] plot ({-2*exp(-2*(\y*\y))},{\y});
		\foreach \x in {0.05,0.1,...,1.0}{
		\fill [opacity=0.020,myred] (2,\ypos) -- ($(7.5,\ypos)+(0,\x)$) -- ($(7.5,\ypos)-(0,\x)$) -- cycle;
		}
	}
	\draw [dashed,draw=myred,domain=-5:5,variable=\y,shift={(imageplane)},smooth,samples=75,line width=0.3mm] plot ({ -2*exp(-2*((\y-4)*(\y-4))) -2*exp(-2*((\y-3.5)*(\y-3.5))) -2*exp(-2*((\y-2.5)*(\y-2.5))) -2*exp(-2*((\y-2)*(\y-2))) -2*exp(-2*((\y-1)*(\y-1))) },{\y});
	\draw [fill=purple,opacity=0.5] (1,0) rectangle (3,5);
\end{tikzpicture}
\caption{Illustration of localisation microscopy with five point sources, imaged by a diffraction-limited system and the resultant intensity distribution on the image plane.}
\label{fig:config}
\end{figure}

Realistic imaging scenarios typically involve more than two point sources or even extended objects.
It has been shown that an extended one-dimensional object much smaller than the Rayleigh limit described only in terms of its centroid and effective radius can be approximated by a two-level quantum system~\cite{ChRaMaBa17}.
Theoretical optimality of certain measurement techniques in estimating this effective radius size has also been established in one and two spatial dimensions~\cite{Tsa17_IOP,Tsa17SPADE,DuKeGu18}.
Order-of-magnitude bounds on the precision of estimating the normalised moments of extended sources smaller the Rayleigh limit have also been obtained~\cite{Zhou17,Tsang18QL}.

In this paper, we provide an analytical lower bound on an unbiased estimator's covariance (mean square error) matrix for localisation microscopy -- simultaneously estimating the locations of $N$ incoherent, weak point sources of unequal but known intensities in one spatial dimension. The bound is provided by the the quantum Fisher information matrix.
For a Gaussian point spread function (PSF), we first describe the light field on the image plane as a classical mixture of coherent states.
We use this to derive the quantum Fisher information matrix analytically. In the limit of the point sources approaching a single point, we find its rank to be no more than two.
As the inverse of the quantum Fisher information matrix lower bounds the covariance matrix, our result implies that no more than two independent parameters can be estimated in localisation microscopy in the limit of arbitrarily small separations.
In this limit, we provide a mathematical explanation for our observation in terms of an approximation of the light field involving only the first two Hermite-Gauss modes.
Finally, we compare performance of conventional direct imaging and the recently proposed SPADE~\cite{TsaNL16} in localisation microscopy with the quantum bounds we obtain. 
In the limit of the point sources approaching a single point, we find the classical Fisher information matrices for both these detection systems to be rank one. Furthermore, in the sub-Rayleigh limit, SPADE does not attain the quantum limit for localisation microscopy. For the subset of parameters where scalings may be optimal, we find SPADE to be short of the quantum limit in absolute precision.

This paper is organised as follows: In Section~\ref{sec:physical_description} we provide a quantum mechanical description of localisation microscopy.
The appropriate framework to study the quantum limits of the localisation problem is quantum estimation theory, the toolbox of which is described in Section~\ref{sec:QET},   leading to the definition of the quantum Fisher information matrix (QFIM).
In Section~\ref{sec:results} we provide an analytic expression of the QFIM for localisation microscopy, our main technical result. We then draw conclusions about its rank and its implications for localisation microscopy. 
We end in Section~\ref{sec:disc} with further insights and discussions about the sinc PSF and the potential of detection systems attaining the quantum limits of localisation microscopy.


\section{Quantum description of localisation microscopy}
\label{sec:physical_description}

We consider localisation microscopy -- the problem of estimating the locations of $N$ incoherent point sources or emitters located in a one-dimensional spatial configuration as in Fig.~\ref{fig:config}.
As we assume them to be weak, such that on average no photons arrive on the image place within a coherence time with probability $(1-\epsilon)$, where $\epsilon \ll 1$ and one photon arrives with probability $\epsilon$.
We also assume the optical field on the image plane to be quasi-monochromatic and paraxial~\cite{TsaNL16}.
The quantum state of this optical field is then
\begin{equation} 
\label{eq:rho_prim}
\rho_{\mathrm{opt}} \approx (1-\epsilon)\rho_{\mathrm{vac}}+\epsilon \rho,
\end{equation}
where we have neglected terms of second and higher orders in $\epsilon$ and $\rho_{\mathrm{vac}}=\ket{\mathrm{vac}} \bra{\mathrm{vac}}$ is the vacuum state and $\rho$ is the one-photon state.
 
 The one-photon density matrix on the object plane is an incoherent mixture of position eigenstates $\rho= \sum_{i=1}^{N} w_i \ket{\chi_{i}} \bra{\chi_{i}} $, where $w_{i}$ are the relative intensities with $\sum_{i=1}^{N} w_{i}=1$.
An imaging system maps $\hat{c}_{x}^{\dagger},$ the creation operator producing one photon in the position $x$ on the object plane, to the corresponding image plane operator $\hat{c}_{i}^\dag$~\cite{LuPi16} 
\begin{equation}\label{eq:creat_oper_x} 
     \hat{c}_{i}^{\dagger}= \int dx \Psi_{\PSF}(x-\chi_{i}) \hat{c}_{x}^{\dagger},
\end{equation}
where $\chi_{i}$ is the position on the source on the object plane and $\psi_{\PSF}(x)$ is the PSF.
On the image plane this becomes
\begin{equation}\label{eq:rho_position}
 \rho=\sum_{i=1}^{N} w_{i} \ket{\psi_{i}} \bra{\psi_{i}},
\end{equation}
where
\begin{equation} \label{eq:psi_pos}
 \ket{\psi_{i}}=\int dx \Psi_{\PSF}(x-\chi_{i}) \ket{x},
\end{equation}
as follows from Eq.~\eqref{eq:creat_oper_x}.

An ideal imaging system with $\Psi_{\PSF}(x) = \delta(x)$ is free of any Rayleigh limit as it transmits all spatial frequencies from the object to the image plane.
In practice, a Gaussian PSF
 \begin{equation}
 \psi_{\PSF}(x)= \frac{1}{(2 \pi \sigma^2)^{1/4}} e^{-\frac{x^{2}}{4\sigma^{2}}}, 
 \end{equation}
with $\sigma= \lambda/(2 \pi \mathrm{NA}),$ where $\mathrm{NA}$ is the numerical aperture of the imaging system is a good approximation for quasimonochromatic paraxial light~\cite{ZhZe06,TsaNL16} and also allows us to obtain analytical results.
For such a PSF, the state of Eq.~\eqref{eq:rho_position} has an intensity distribution of the form illustrated in Fig.~\ref{fig:config}.
For a Gaussian PSF, the $\ket{\psi_{i}}$ can be expanded in the Hermite-Gauss (HG) basis as (See Appendix \ref{app:HG_basis})
\begin{equation} 
 \label{eq:psi_HG}
|\psi_{i} \rangle =\sum_{k=0}^{\infty} \frac{\alpha_{i}^{k}}{\sqrt{k!}} e^{-\alpha_{i}^{2}/2}  |\phi_{k} \rangle \equiv  \ket{\alpha_{i}},
\end{equation}
where $\ket{\phi_{k}}$ are the HG modes%
\footnote{Unlike the conventional quantum optical coherent states which reside in the phase space of the electromagnetic field, our coherent states reside in physical space on the image plane.
This mathematical form was also identified by~\citet{DuKeGu18} but only used for numerical calculations.
}
This has the same mathematical form as the coherent states, produced by the displacement operator \( \mathcal{D}(\alpha_i) = e^{\alpha_i\op{a}^{\dagger}+\alpha_i^*\op{a}} \)~\cite{kok_lovett_2010} acting on the ground state of the harmonic oscillator with $\alpha_{i}=\chi_{i}/2\sigma \in \mathbb{R}$ the dimensionless positions of the sources.
Thus the one-photon state on the image plane is
 \begin{equation}
 \label{eq:rho_sum_coh}
 \rho \equiv \rho_{\bm{\alpha}} = \sum_{i=1}^{N} w_{i} \ket{\alpha_{i}} \bra{\alpha_{i}},
 \end{equation}
 a classical mixture of coherent states in the HG basis.

The above is a quantum optical rendition of localisation microscopy---a classical optics problem.
It enables us to harness the mathematical formalism associated with coherent states and provides a basis that spans the space of the quantum state as well as its derivative.
The latter is an essential ingredient of deriving the quantum Fisher information matrix analytically in Section~\ref{sec:QFI_Analyt}.
We also hope that this description will provide insights into the quantum limits to localisation microscopy in the presence of shot noise and assist in designing detection systems that attain these quantum limits.


\section{Quantum Estimation Theory}
\label{sec:QET}


Localisation has long been treated as an estimation problem with the unknown locations of the sources $\bm{\chi} \equiv \{ \chi_{i} \},~i=1,\dots,N$ being the parameters to be estimated~\cite{Ober2004,ChSaO16}.
In our formulation, the limits to the localisation of the point sources are the same as estimating the amplitudes $\bm{\alpha} \equiv \{ \alpha_{i} \},~i=1,\dots,N$ of the coherent states in Eq.~\eqref{eq:rho_sum_coh}. 
Let these estimates be $\tilde{\bm{\alpha}} \equiv \{ \tilde{\alpha}_{i} \}$.
The precision of our estimate is then given by the covariance (or mean square error) matrix defined as
\begin{equation} 
\label{eq:cov}
 \operatorname{Cov}[\bm{\alpha}]  = \sum_z p(z|\bm{\alpha}) (\bm{\alpha} - \tilde{\bm{\alpha}})^T(\bm{\alpha} - \tilde{\bm{\alpha}}),
\end{equation}
where $p(z|\bm{\alpha})$ is the probability distribution of the collected data labelled by, for instance, the pixel $z$ on the image plane.
$\operatorname{Cov}[\bm{\alpha}]$ is a positive symmetric matrix whose $i-$th diagonal element denotes the variance of an estimator of $\alpha_{i}$ given the data collected.
The $(i,j)$-th off-diagonal element denote the covariance in the estimation of $\alpha_{i}$ and $\alpha_{j}$. 

Given the data collected, the maximum amount of information that can be extracted from it to obtain the most precise estimate of the locations is given by the Cram\'er-Rao bound (CRB)~\cite{ThomasCover}. 
For an unbiased estimator, this bound is given by
\begin{equation} \label{eq:CCRB}
 \operatorname{Cov}[\bm{\alpha}] \geq \frac{1}{M\epsilon} \left[\mathcal{C}(\rho_{\bm{\alpha}}, \mathrm{\Pi}_{z})\right]^{-1},
\end{equation}
 where $M$ is the number of coherence times over which the data is collected, making $M\epsilon$ the total photon count.
This inequality is saturable but generally only in the asymptotical limit of many repetitions~\cite{VanTrees}. 
$\mathcal{C}(\rho_{\bm{\alpha}}, \mathrm{\Pi}_{z})$ is the classical Fisher information matrix (CFIM) whose elements are given by~\cite{Par09} 
\begin{equation} \label{eq:CFIM}
[\mathcal{C}(\rho_{\bm{\alpha}}, \mathrm{\Pi}_{z})]_{\mu \nu} =
\sum\limits_z \frac{1}{p(z|\bm{\alpha} )} \frac{\partial p(z|\bm{\alpha})}{\partial \alpha_{\mu}} \frac{\partial p(z|\bm{\alpha})}{\partial \alpha_{\nu}}.
\end{equation}
The probability distribution $p(z|\bm{\alpha}) = \mathrm{Tr}\left( \rho_{\bm{\alpha}} \mathrm{\Pi}_{z} \right)$ results from detecting the light on the image plane using a specific detection system $\mathrm{\Pi}_{z}$.
Fluorescence microscopy typically employs intensity detectors $\mathrm{\Pi}_{z} = \{\proj{n}_z\}, n = 0, 1, \cdots,$ at each pixel $x,$ known as direct imaging.
It is then evident that the CFIM depends on the detection system used, and not surprising that it determines the amount of information that can be extracted from the light field at the image plane. 

To identify the quantum limit on the precision of localisation microscopy, the CFIM must be maximised over all possible physically allowed detection systems.
This set is given by positive operator-valued measures (POVMs)~\cite{NielsenChuang} and the maximisation is bounded as~\cite{Holevo,Helstrom}.
\begin{equation} 
 \label{eq:maxCCRB}
 \max _{\{\Pi_z\}} ~ 
  \mathcal{C}(\rho_{\bm{\alpha}}, \mathrm{\Pi}_{z}) \leq \mathcal{Q}(\rho_{\bm{\alpha}}),
\end{equation}
by the quantum Fisher information matrix (QFIM). 
Its matrix elements are given by
\begin{equation}\label{eq:QFIM}
[\mathcal{Q}(\rho_{\bm{\alpha}})]_{\mu \nu }=\mathrm{Tr} \left[ \rho_{\bm{\alpha}} \frac{L^{\mu}L^{\nu}+L^{\nu}L^{\mu}}{2}  \right],
\end{equation}
with $L_{\mu}$ being the symmetric logarithmic derivative (SLD) corresponding to the parameter $\alpha_{\mu}$.
The SLD is determined by the Lyapunov equation 
\begin{equation}\label{eq:Lyap_def}
2\frac{\partial \rho_{\bm{\alpha}}}{\partial \alpha_{\mu}} = ( \rho_{\bm{\alpha}}L^{\mu} +L^{\mu} \rho_{\bm{\alpha}}).
\end{equation}

The quantum limit to localisation microscopy is thus given by
\begin{equation} \label{eq:QCRB}
 \operatorname{Cov}[\bm{\alpha}] \geq \frac{1}{M\epsilon} \left[\mathcal{C}(\rho_{\bm{\alpha}}, \mathrm{\Pi}_{z})\right]^{-1} \geq \frac{1}{M\epsilon} \left[\mathcal{Q}(\rho_{\bm{\alpha}})\right]^{-1}.
\end{equation}
The QFIM depends only on the light field on the image plane and determines the maximum amount of information that can be extracted from it using detection systems allowed by quantum mechanics.
Deriving an analytical expression for $\mathcal{Q}(\rho_{\bm{\alpha}})$ for state in Eq.~\eqref{eq:rho_sum_coh} is our main result, which we present in the next section.

A practical issue following the identification of the quantum limit is its attainability.
For cases where a single parameter is unknown then a measurement can be found to satisfy the equality of Eq.~\eqref{eq:maxCCRB}, which involves projecting onto the eigenstates of the SLD~\cite{CaBr94,Par09}. However this strategy does not generalise to multiple parameters, as in localisation microscopy, in general.

For multi-parameter estimation the attainability is tantamount to saturating the second inequality in Eq.~\eqref{eq:QCRB}. A necessary condition for the saturability of any scalar form of Eq.~\eqref{eq:QCRB} is the satisfaction of \emph{weak commutativity}~\cite{matsumoto_new_2002,RaJRaf16}
 \begin{equation} \label{eq:weak_comm}
	 \mathrm{Tr}\left(\rho_{\bm{\alpha}} [L^{\mu},L^{\nu}]\right)= 2\mathrm{Tr}\left(\mathrm{Im}\left( \rho_{\bm{\alpha}} L^{\mu}L^{\nu} \right)\right) = 0.
 \end{equation}
Moreover, through the quantum theory of asymptotic normality~\cite{kahn_local_2009}, this condition becomes sufficient with the application of collective measurements over multiple copies of $\rho_{\bm{\alpha}}$ ~\cite{matsumoto_new_2002,RaJRaf16}.

Any scalar function of the covariances can be bounded by the inverse of QFIM with the lower bound following from the spectral decomposition of QFIM. To that end, calculating the eigenvalues of the QFIM and their scaling is of importance for the multi-parameter estimation.
For localisation microscopy, $\rho_{\bm{\alpha}}$ as in Eq.~\eqref{eq:rho_sum_coh} as well as its derivative are real matrices. Thereby, $L^{\mu},L^{\nu}$ are also real and the above condition is always satisfied%
\footnote{Since the density matrix and its derivatives are real-valued in the orthonormal \( \{ \ket{\phi_k} \} \) basis, Eq.~\eqref{eq:Lyap_def} is a system of equations with real coefficients.
Hence $L^{\nu}$ must be real as well, and so \( \rho_{\bm{\alpha}} L^{\mu}L^{\nu} \) is real-valued. We thank Ben Wang for bringing this to our attention}.
The quantum limit for localisation microscopy is therefore attainable, at least in principle, although collective measurements over multiple copies~\cite{matsumoto_new_2002,RaJRaf16} of the light field on the image plane may be required.

Alternative parameterisations of the system---where the new parameters $\bm{\alpha}'$ are functions of the old parameters $\bm{\alpha}$---can be dealt with by a transformation of the QFIM.
Given the transformation matrix $B$ with elements $B_{ij}=\partial \alpha_{i}/ \partial \alpha'_{j}$, the QFIM of the transformed parameters is~\cite{Par09}
 \begin{equation} \label{eq:QFI_trans}
	 \mathcal{Q}'=B \, \mathcal{Q} \, B^{T},
 \end{equation}
provided the transformation is non-singular. This can be used to recast our results in terms of, for instance, the moments of the point source distribution.


\section{Results} \label{sec:results}

We now present our main result -- the analytical expression of the QFIM for localisation microscopy.
This expression allows us to conclude that the QFIM is a rank two matrix as $\alpha_i \rightarrow 0$.
Eq.~\eqref{eq:QCRB} then implies that the eigenvalues of $\operatorname{Cov}[\bm{\alpha}]$ remains finite for no more than two independent parameters.
Thus, no more than two independent parameters can be estimated from the entire set $\bm{\alpha}$ as $\bm{\alpha} \rightarrow 0$. 

We lack a fully satisfactory physical explanation for this restriction on the number of estimable parameters, but provide an explanation involving only the first two Hermite-Gauss modes for $\alpha_i \ll 1$.

\subsection{Analytical expression of QFIM} \label{sec:QFI_Analyt}

The state in Eq.~\eqref{eq:rho_sum_coh} can be expressed in the basis of $ \{ \ket{\alpha_i}, \hat{a}^{\dagger}\ket{\alpha_{i}} \} $ as
\begin{equation} \label{eq:rho_lyap}
	\rho_{\bm{\alpha}}   = A \begin{pmatrix} D_{\bm{w}} & 0 \\ 0 & 0 \end{pmatrix} A^{\dagger} \equiv A \rho_A A^{\dagger} 
\end{equation}
where
\begin{equation}
	{A} = \begin{pmatrix} \ket{\alpha_1} & \ket{\alpha_2} & \cdots & \ket{\alpha_N} & \op{a}^{\dagger} \ket{\alpha_1} & \cdots & \op{a}^{\dagger} \ket{\alpha_N} \end{pmatrix}
\end{equation}
and $ D_{\bm{w}} = \diag \left( w_1 , w_2 , \cdots , w_N \right) $ denotes a diagonal matrix.
Although the basis used in Eq.~\eqref{eq:rho_lyap} is non-orthogonal this representation can still be used to evaluate the QFIM~\cite{GeTu19}.
The coherent states  $\{ \ket{\alpha_{i}} \}$ are linearly independent and span the support of the state in Eq.~\eqref{eq:rho_sum_coh}.
The support of the derivative is spanned by $\{\ket{\alpha_{i}} \}$ and $\{ \hat{a}^{\dagger}\ket{\alpha_{i}} \}$,  which are also linearly independent.

The Grammian matrix
\begin{equation}
	\Upsilon = A^{\dagger} A,
\end{equation}
whose elements consist of the scalar products between the basis vectors \( \braket{\alpha_j | \alpha_k} \), \( \braket{\alpha_j | \op{a}^{\dagger} | \alpha_k} \), \( \braket{\alpha_j | \op{a} | \alpha_k} \), and \( \braket{ \alpha_j | \op{a} \op{a}^{\dagger} | \alpha_k} \)
is in block form,
\begin{equation}
\Upsilon =	\begin{pmatrix}
		\Upsilon_{\alpha \alpha} &  \Upsilon_{\alpha d}  \\
		\Upsilon_{d \alpha} & \Upsilon_{d d}
	\end{pmatrix},
\end{equation}
where 
\begin{equation}
\begin{aligned}
 \left(	\Upsilon_{\alpha \alpha} \right)_{ij} &= \braket{\alpha_i | \alpha_j} = e^{ - (\alpha_i-\alpha_j)^2/2 }, \\
\left( 	\Upsilon_{\alpha d } \right)_{ij}= \Upsilon_{d \alpha}^{\dagger} &= \braket{\alpha_i | \op{a}^{\dagger} | \alpha_j} = \alpha_i e^{ -(\alpha_i-\alpha_j)^2/2}=D_{\bm{\alpha}} \Upsilon_{\alpha \alpha}, \\
\left( \Upsilon_{d d} \right)_{ij}&= \braket{\alpha_i | \op{a} \op{a}^{\dagger} | \alpha_j} = (\alpha_i \alpha_j + 1) e^{ -(\alpha_i-\alpha_j)^2/2 }\\ &=D_{\bm{\alpha}} \Upsilon_{\alpha \alpha} D_{\bm{\alpha}} + \Upsilon_{\alpha \alpha},
\end{aligned}
\label{eq:upsilon_blocks}
\end{equation}
and  $D_{\bm{\alpha}} = \diag \left( \alpha_1 , \alpha_2 , \cdots , \alpha_n \right).$

Since \( \partial_{\alpha} \ket{\alpha} = (\op{a}^{\dagger} - \alpha) \ket{\alpha} \) for real \( \alpha \), 
the derivative of the quantum state is
\begin{equation} \label{eq:rho_der}
	\partial_j \rho_{\bm{\alpha}} = A  w_{j}\begin{pmatrix} -2\alpha_{j} E_j & E_j \\ E_j & 0 \end{pmatrix}  A^{T} \equiv A (\partial_j\rho)_A A^{\dagger},
\end{equation} 
where \( \partial_j \) denotes the derivative with respect to \( \alpha_j \) and \( (E_j)_{kl} = \delta_{jk} \delta_{jl} \).
Similarly, the SLD $L_{A}^i$ can be written in the generic form 
\begin{equation}
L^{j} = A L_{A}^{j} A^{T} = A \begin{pmatrix}
L_{\alpha \alpha}^{j} & L_{\alpha d}^{j}\\
L_{d\alpha}^{j} & L_{dd}^{j}
\end{pmatrix} A^{T}
\end{equation}
where $L_{\alpha\alpha}^j$ corresponds to the elements $\bra{\alpha_{i}} L^j \ket{\alpha_{j}}$, $L_{\alpha d}^j$ to $ \bra{\alpha_{j} } L^j \hat{a}^{\dagger} \ket{\alpha_{i}} $ etc.
The Lyapunov equation Eq.~\eqref{eq:Lyap_def} can be now rewritten as
\begin{equation} \label{eq:Lyap_tof}
	2\partial_{j} \rho_{A} = \rho_{A} \Upsilon L_{A}^{j}+L_{A}^{j} \Upsilon \rho_{A},
\end{equation}
and the QFIM elements from Eq.~\eqref{eq:QFIM} as
\begin{equation}
[\mathcal{Q}(\rho_{\bm{\alpha}})]_{jk}= \Trace{ \partial_{j} \rho \, L^{k}}=\Trace{ \partial_{j} \rho_{A} \Upsilon \, L^{k}_{A} \Upsilon}.
\label{eq:qfi_elements_trace}
\end{equation}

Using the Tracy-Singh block kronecker product $\TracySingh$ and the block column "$\mathrm{vecb}$" operator~\cite{KoNeWa91} defined as
\begin{equation}
\vecb{L_{A}^{j}} =\begin{bmatrix}
\Vect{ L_{\alpha \alpha}^{j} }\\
\Vect{ L_{d \alpha}^{j} } \\
\Vect{L_{\alpha d}^{j} }\\
\Vect{L_{dd}^{j}}
\end{bmatrix},
\end{equation}
where $\Vect{X}=\mathrm{vec}(X)$ is the column vectorisation of a matrix and $\Vectr{X}$ its transpose.
Eq.~\eqref{eq:Lyap_tof} can be blockwise vectorised to
\begin{equation}\label{eq:Lyap_vecbc}
 2 \vecb{ \partial_{j} \rho_{A} } =\left( \mathbb{I} \TracySingh (\rho_{A} \Upsilon ) + (\rho_{A} \Upsilon ) \TracySingh \mathbb{I} \right) \vecb{L_{A}^{j}} 
\end{equation}
with \( \mathbb{I} \) being the identity matrix.
Using the matrix identity~\cite{KoNeWa91} 
\begin{equation}
\Trace{ A^{T} B C D^{T} }=\vecb{A^{T}}^{T}(D \TracySingh B) \vecb{C},
\end{equation}
the QFIM elements from Eq.~\eqref{eq:qfi_elements_trace} can be re-expressed as
\begin{equation}
\begin{aligned}
[ \mathcal{Q}(\rho_{\bm{\alpha}})]_{ij}&= \vecb{\partial_{i} \rho_{A}}^{T} (\Upsilon \TracySingh \Upsilon) \vecb{L_{A}^{j}} \\
&= w_{i} \begin{bmatrix}
-2 \alpha_{i} \Vectr{E_{i}} & \Vectr{E_{i}} &  \Vectr{E_{i}} & 0
\end{bmatrix} 
\begin{bmatrix}
\Vect{ \Gamma_{\alpha \alpha}^{j} }\\
\Vect{ \Gamma_{d \alpha}^{j} } \\
\Vect{\Gamma_{\alpha d}^{j}}\\
\Vect{\Gamma_{dd}^{j}}
\end{bmatrix},
\end{aligned}
\label{eq:qfie_gam}
\end{equation}
where we have defined 
\begin{equation} \label{eq:Gamma_def}
(\Upsilon \TracySingh \Upsilon) \vecb{L^j_{A}} 
= \vecb{\Gamma^{j}} 
= \begin{bmatrix}
\Vect{ \Gamma_{\alpha \alpha}^{j} }\\
\Vect{ \Gamma_{d \alpha}^{j} } \\
\Vect{\Gamma_{\alpha d}^{j}}\\
\Vect{\Gamma_{dd}^{j}},
\end{bmatrix}
\end{equation}
which is the outstanding quantity to be determined.

We now recast Eq.\eqref{eq:Lyap_vecbc} and \eqref{eq:qfie_gam} as
\begin{equation} \label{eq:Lyap_gamma_fin}
\begin{aligned}
2 \vecb{ \partial_{i} \rho_{A} }& =\left( \Upsilon^{-1} \TracySingh \rho_{A} + \rho_{A} \TracySingh \Upsilon^{-1} \right) (\Upsilon \TracySingh \Upsilon) \vecb{L_{A}^{i}} \\
&=\left( \Upsilon^{-1} \TracySingh \rho_{A} + \rho_{A} \TracySingh \Upsilon^{-1} \right) \vecb{\Gamma^{i}}.
\end{aligned}
\end{equation}
Putting it all together, we obtain
\begin{equation} \label{eq:Lyap_matrix_form}
\begin{bmatrix}
-4 w_{i} \alpha_{i} \Vect{E_{i}}\\
2 w_{i} \Vect{E_{i}} \\
2 w_{i} \Vect{E_{i}}\\
0
\end{bmatrix} = \begin{bmatrix}
~ & ~ &~ & 0 \\
~ & \mathbb{A} & ~ &0 \\
~ & ~ & ~ &0 \\
0& 0& 0& 0
\end{bmatrix} \, \begin{bmatrix}
\Vect{ \Gamma_{\alpha\alpha}^{j} }\\
\Vect{ \Gamma_{d\alpha}^{j} } \\
\Vect{\Gamma_{\alpha d}^{j}}\\
\Vect{\Gamma_{dd}^{j}}
\end{bmatrix},
\end{equation}
\vspace{1ex}
\noindent where
\begin{equation} 
\label{eq:A_matrix}
\mathbb{A}=\begin{bmatrix}
D_{\bm{w}} \otimes \upsilon_{\alpha\alpha}+\upsilon_{\alpha\alpha} \otimes D_{\bm{w}} & D_{\bm{w}} \otimes \upsilon_{\alpha d} & \upsilon_{\alpha d} \otimes D_{\bm{w}}  \\
 D_{\bm{w}} \otimes \upsilon_{d\alpha} & D_{\bm{w}} \otimes \upsilon_{dd} & 0 \\
\upsilon_{d\alpha} \otimes D_{\bm{w}} & 0 & \upsilon_{dd} \otimes D_{\bm{w}} 
\end{bmatrix},
\end{equation}
and $\{ \upsilon_{\alpha\alpha}, \upsilon_{\alpha d}, \upsilon_{d\alpha}, \upsilon_{dd} \}$ defines the inverse of $\Upsilon$ via
\begin{equation}
\Upsilon^{-1} = 
\begin{bmatrix}
\upsilon_{\alpha\alpha} & \upsilon_{\alpha d} \\
\upsilon_{d\alpha} & \upsilon_{dd}
\end{bmatrix}.
\end{equation}
Note that the inverse $\Upsilon^{-1}$ always exists since $\Upsilon$ is the Grammian matrix of linearly independent vectors.
The elements of $\Upsilon^{-1}$ can be found using the formula of blockwise inversion (See Appendix \ref{app:Analyt}).

Noticing that $\Vect{\Gamma_{dd}^{j}}$ does not contribute in Eq.~\eqref{eq:qfie_gam}, Eq.~\eqref{eq:Lyap_matrix_form} can be reduced to
\begin{equation} 
\label{eq:Lyap_reduced}
\begin{bmatrix}
-4 w_{i} \alpha_{i} \Vect{E_{i}}\\
2 w_{i} \Vect{E_{i}} \\
2 w_{i} \Vect{E_{i}}
\end{bmatrix} 
= \mathbb{A} \begin{bmatrix}
\Vect{ \Gamma_{\alpha\alpha}^{j} }\\
\Vect{ \Gamma_{d\alpha}^{j} } \\
\Vect{\Gamma_{\alpha d}^{j}}
\end{bmatrix}
\end{equation}
where
$\mathbb{A}$ is a $3N^{2} \times 3N^{2}$ invertible matrix unless $\alpha_{i} =\alpha_{j}$ for some $i,j$, which is a singular case for which the rank of the density matrix reduces. 
Hence the unique solution to Eq.~\eqref{eq:Lyap_reduced} is 
\begin{equation}
\begin{bmatrix}
\Vect{ \Gamma_{\alpha\alpha}^{j} }\\
\Vect{ \Gamma_{d\alpha}^{j} } \\
\Vect{\Gamma_{\alpha d}^{j}}
\end{bmatrix}= \mathbb{A}^{-1} \begin{bmatrix}
-4 w_{j} \alpha_{j} \Vect{E_{j}}\\
2 w_{j} \Vect{E_{j}} \\
2 w_{j} \Vect{E_{j}}
\end{bmatrix},
\label{eq:gamma_with_inverse}
\end{equation}
where the block matrices that compose the $\mathbb{A}^{-1}$ can be found by using the formulas for blockwise inversion (See Appendix~\ref{app:Analyt}).

Substituting Eq.~\eqref{eq:gamma_with_inverse} into Eq.~\eqref{eq:qfie_gam} gives us the QFIM elements
\begin{widetext}
\begin{equation} \label{eq:qfie_final}
\begin{aligned}
[\mathcal{Q}(\rho_{\bm{\alpha}})]_{ij}
&= 2 w_{i} w_{j} 
\Vectr{E_{i}} \left[
\mathbb{I} \otimes \Upsilon_{d\alpha} \Upsilon_{\alpha\alpha}^{-1} + \Upsilon_{d\alpha} \Upsilon_{\alpha\alpha}^{-1} \otimes \mathbb{I} - 2\alpha_i \mathbb{I} \otimes \mathbb{I}
\right] S^{-1} \left[
\mathbb{I} \otimes \Upsilon_{\alpha\alpha}^{-1} \Upsilon_{\alpha d} + \Upsilon_{\alpha\alpha}^{-1} \Upsilon_{\alpha d} \otimes \mathbb{I} - 2\alpha_j \mathbb{I} \otimes \mathbb{I}
\right] \Vect{E_j} \\
&\mkern16mu +
4 w_i \delta_{ij}
\left[ 1+\alpha_i^2 - (\Upsilon_{\alpha\alpha}D_{\alpha}\Upsilon_{\alpha\alpha}^{-1}D_{\alpha}\Upsilon_{\alpha\alpha})_{ij} \right]
\end{aligned}
\end{equation}
\end{widetext}
where \( S^{-1} = \left( \Upsilon_{\alpha\alpha}^{-1} \otimes D_{\bm{w}} + D_{\bm{w}} \otimes \Upsilon_{\alpha\alpha}^{-1} \right)^{-1} \) is an $N^{2} \times N^{2}$ matrix and $\Upsilon_{\alpha \alpha}$ is the inverse of the submatrix of $\Upsilon$ which exists, as it is the Grammian matrix of linearly independent vectors $\{ \ket{\alpha_{i}} \}$. 
Eq.~\eqref{eq:qfie_final} is an analytic expression for the QFIM elements for localisation microscopy and our main result.

\begin{figure}[htb]
\centering
 \includegraphics[scale=0.36]{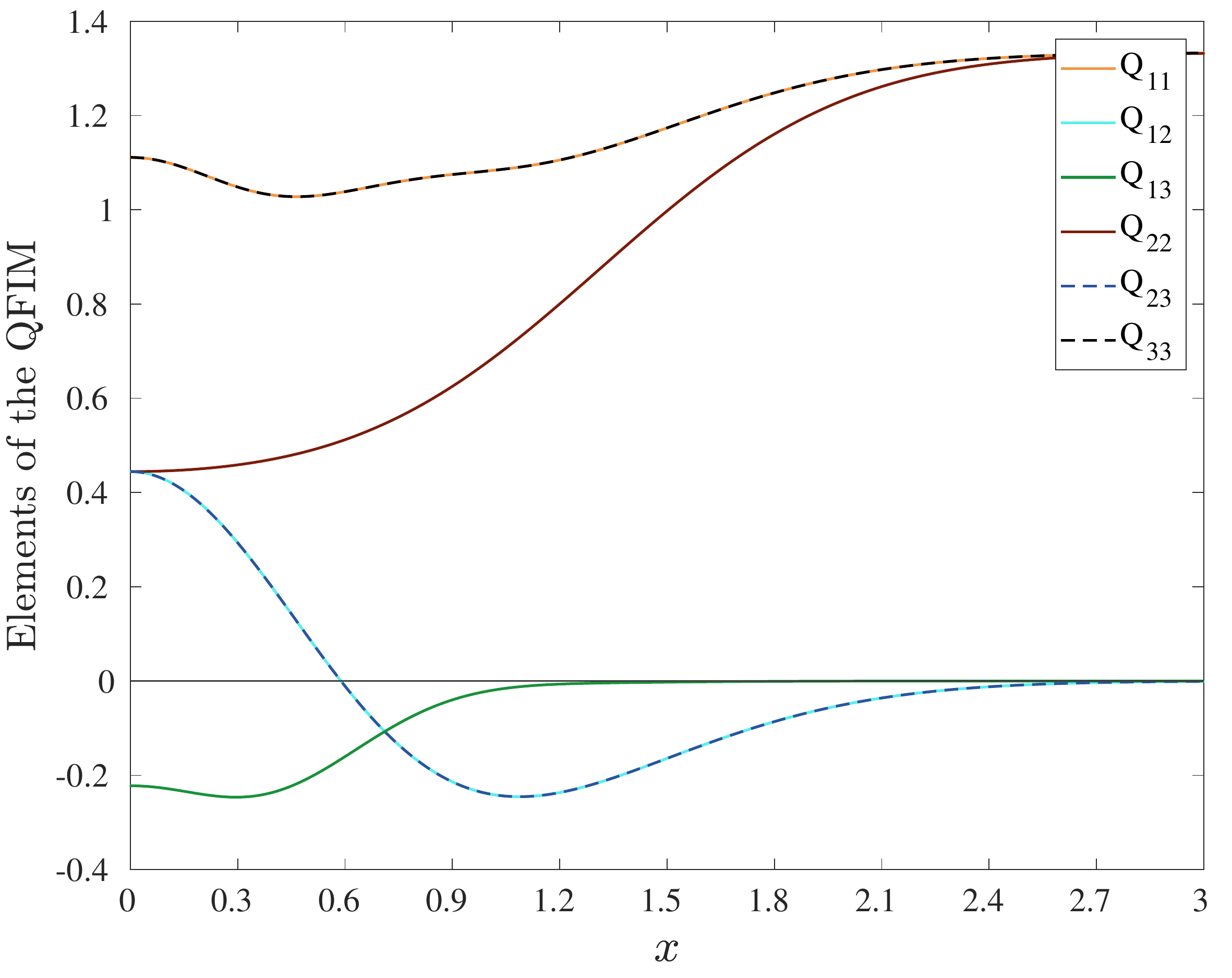}
 \label{fig:qfi_inv}
 \caption{\small Diagonal and off diagonal elements of the QFIM for the case of 3 sources with equal intensities. The sources are separating from each other at equal distances, $(\alpha_{1},\alpha_{2},\alpha_{3})=(x,2x,3x)$. The element $\mathcal{Q}_{12}$ and $\mathcal{Q}_{23}$ elements are equal, as are the $\mathcal{Q}_{11}$ and $\mathcal{Q}_{33}$ elements.}
\label{fig:QFI_elem}
\end{figure}

Fig.~\ref{fig:QFI_elem} shows the elements of the QFIM for the localisation microscopy of three point sources.
We choose them to be equidistant, that is, $(\alpha_{1},\alpha_{2},\alpha_{3})=(x,2x,3x)$ and $w_1 = w_2 = w_3 = 1/3$ for illustration purposes.
Note the non-zero off-diagonal elements evidencing correlations in the precision around and below the Rayleigh limit of $x \sim 1.$ 

\begin{figure}[htb] 
\centering
\includegraphics[scale=0.355]{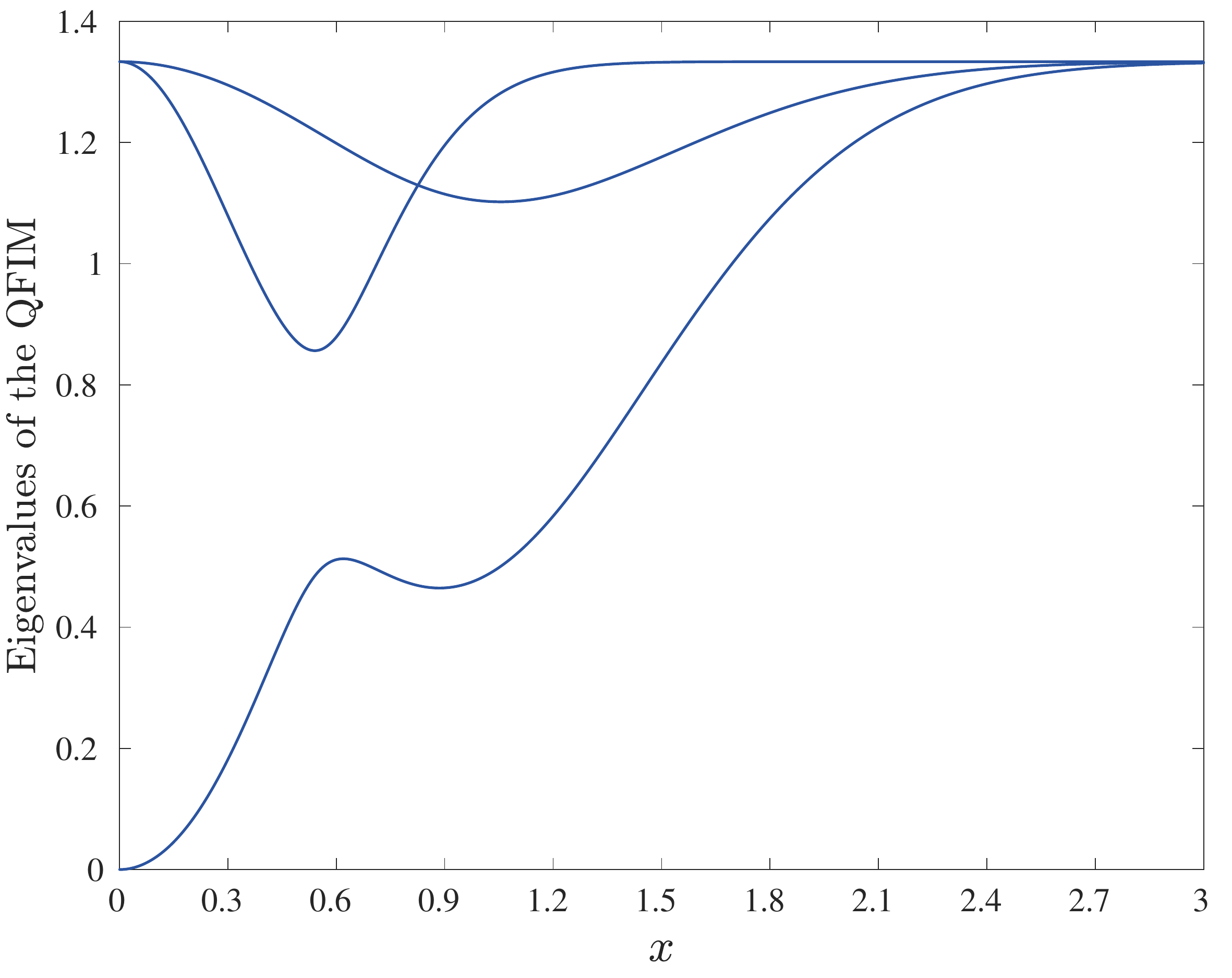}
\caption{\small The eigenvalues of the QFIM matrix for 3 sources with equal intensities. The sources are separating from each other at equal distances, $(\alpha_{1},\alpha_{2},\alpha_{3})=(x,2x,3x)$.}
\label{fig:EIGENV} 
\end{figure}

While the diagonal elements are all non-vanishing, more crucially as $x \rightarrow 0$ the diagonal and off-diagonal elements combine to make the QFIM singular.
This is revealed by a closer analysis of the QFIM matrix as in Fig.~\ref{fig:EIGENV} which shows that only two of its eigenvalues remain non-zero as the sources approach each other. 
This is in spite of all the diagonals elements of the QFIM remaining non-zero even as $x \rightarrow 0$, as Fig.~\ref{fig:QFI_elem} shows.

This behaviour of only two non-zero eigenvalues also holds for other values of $N$.
 We have explicitly checked this for $N=4,\dots ,10$ as well as when the sources are not equally spaced.
In Fig.~\ref{fig:eig45} in Appendix \ref{app:Analyt} we plot the eigenvalues of the QFIM for $N=4,5$ as further examples. 
In the case of different relative intensities the results are the same except of the limiting case of one extremely bright source $w_{j} \gg 1, w_{i\neq j} \ll 1$, where the rank of the QFIM is approximately one (Fig.~\ref{fig:qfi_unequal_weights} in Appendix~\ref{app:Analyt}).

Since the QFIM has rank two as $x \rightarrow 0,$ its inverse is ill-defined except on a two-dimensional subspace.
This implies that the $N \times N$ covariance matrix for localisation microscopy, as per Eq.~\eqref{eq:QCRB}, will also be unbounded except on a two-dimensional subspace.
Thus, no more than two independent parameters can be estimated in localisation microscopy as the point sources approach each other.

In other words, the rank-deficient nature of the QFIM shows that a form of the Rayleigh limit resurfaces for any $N>2$.
This had been suggested by previous works based on order-of-magnitude bounds for the diagonal elements on the CFIM~\cite{Zhou17} or uppers bounds on the diagonal elements of the QFIM~\cite{Tsang18QL}.
Our analytical expression for the full QFIM---its diagonal and off-diagonal elements for any $N$---shows that this rank two behaviour is truly quantum mechanical in origin.
Furthermore, knowing the full QFIM matrix allows us to uncover the nature in which $N-2$ of the eigenvalues approach zero. We return to the behaviour in which this rank deficiency or Rayleigh limit emerges in Sec.~\ref{sec:disc}.

\subsection{Why rank two?}
\label{sec:why_rank2}

We now provide an explanation for the rank deficiency of the QFIM in the regime of small separations which can be seen as the re-emergence of the Rayleigh limit.
To that end, we expresses the state in Eq.~\eqref{eq:rho_sum_coh} in terms of the real-valued displacement operator $\mathcal{D}(\alpha_{i})= e^{\alpha_{i} \hat{c}_i^{\dagger}-\alpha_{i} \hat{c}_i}$ as 
\begin{equation} \label{eq:rho_channel}
    \rho=\sum_{i} \sqrt{w_{i}}~\mathcal{D}(\alpha_{i}) \ket{0} \bra{0} \sqrt{w_{i}}~\mathcal{D}^{\dagger}(\alpha_{i}).
\end{equation}
In the limit of very small separations ($\alpha_i \ll 1$), the displacements are approximately
\begin{equation} 
    \mathcal{D}(\alpha_{i})=\mathbb{I}+\alpha_{i} \left( \hat{a}^{\dagger}-\hat{a} \right) + \mathcal{O}(\alpha_i^2)  ,
\end{equation}
where $\mathbb{I}$ is the identity operator and the displacement $\alpha_i$ is real.
Up to the second order in $\alpha_i$, the normalised quantum state of the light field on the image place is then
\begin{equation}\label{eq:rho_qubit}
 \rho_{\bm{\alpha}}^{(2)}=
 \begin{pmatrix}
    1-\mathcal{C}_{2} &  \mathcal{C}_{1} \\
    \mathcal{C}_{1} & \mathcal{C}_{2} \\
 \end{pmatrix},
\end{equation}
where \( \mathcal{C}_i \) are the first two moments
\begin{equation}
 	\mathcal{C}_{1}=\sum_{i=1}^{N} w_{i} \alpha_{i}, \quad \mathcal{C}_{2}=\sum_{i=1}^{N} w_{i} \alpha_{i}^{2}.
\end{equation}

Eq.~\eqref{eq:rho_qubit} describes the state of two-level quantum system---the two levels being the first two HG modes.
A similar approximation which described the state relative to a PSF centred at a fixed reference point was used in Ref.~\cite{ChRaMaBa17} to estimate the centroid and the effective radius of a distribution of incoherent point sources. 
We now consider the more general problem of estimating the location of $N$ point sources.

The QFIM for $\bm{\alpha}$ (See Appendix~\ref{app:eigen}) is
\begin{equation} \label{eq:QFIm_qubit}
\mathcal{Q}\left(\rho_{\bm{\alpha}}^{(2)}\right) \equiv \mathscr{Q}=\frac{1}{\mathcal{A}} \left( \mathbf{I} \quad \bm{\alpha}   \right) 
\mathcal{M}
\begin{pmatrix}
\mathbf{I}^{T}   \\
\bm{\alpha}^{T}  
\end{pmatrix},
\end{equation}
with
\begin{equation}\label{eq:qfi_qubit_coef}
    \mathcal{M} = \begin{bmatrix}
        M_{11} & M_{12} \\ M_{21} & M_{22}
    \end{bmatrix},
\end{equation}
where  $M_{11}= (\mathcal{C}_{2}-1)\mathcal{C}_{2}$,
$M_{12}=M_{21}=\mathcal{C}_{1}( 1-2  \mathcal{C}_{2})$,
$M_{22}=4 \mathcal{C}_{1}^{2} -1$,
$\mathcal{A}=\left( \mathcal{C}_{2}-1\right)\mathcal{C}_{2}+\mathcal{C}_{1}^{2}$,
and  $\bm{I}=\left( 1 \; 1 \dots 1 \right)^{T}$.

The QFIM $\mathscr{Q}$ is an $ N \times N $ matrix, which is a product of three matrices of dimensions $N \times 2$, $2 \times2$ and $2 \times N$.
Since $\rank (AB) \leq \min \left\{ \rank (A), \rank (B) \right\}$, and the matrix $\mathcal{M}$ has rank 2, the QFIM $\mathscr{Q}$ has rank no more than two.
Although a two-level quantum system has the potential of estimating three real parameters, localisation microscopy in this limit can estimate only two as the two-level system possesses a real density matrix%
\footnote{As the localisation parameters $\bm{\alpha}$ are real, $\mathrm{Tr}\left( \rho_{\bm{\alpha}}^{(2)} \, \sigma_{y} \right)=2 \, \mathrm{Im}\left( \mathcal{C}_{1}\right) =0,$ where $\sigma_{y} $ is the Pauli $Y$ matrix.}
This is another way of arguing that as the point sources get closer, the light field on the image plane has enough information to estimate only two parameters.
A physical reason for this observation would be highly desirable.

\section{Discussion}
\label{sec:disc}

\begin{figure}[!b] 
\begin{minipage}[b]{0.49\textwidth}
\includegraphics[scale=0.38]{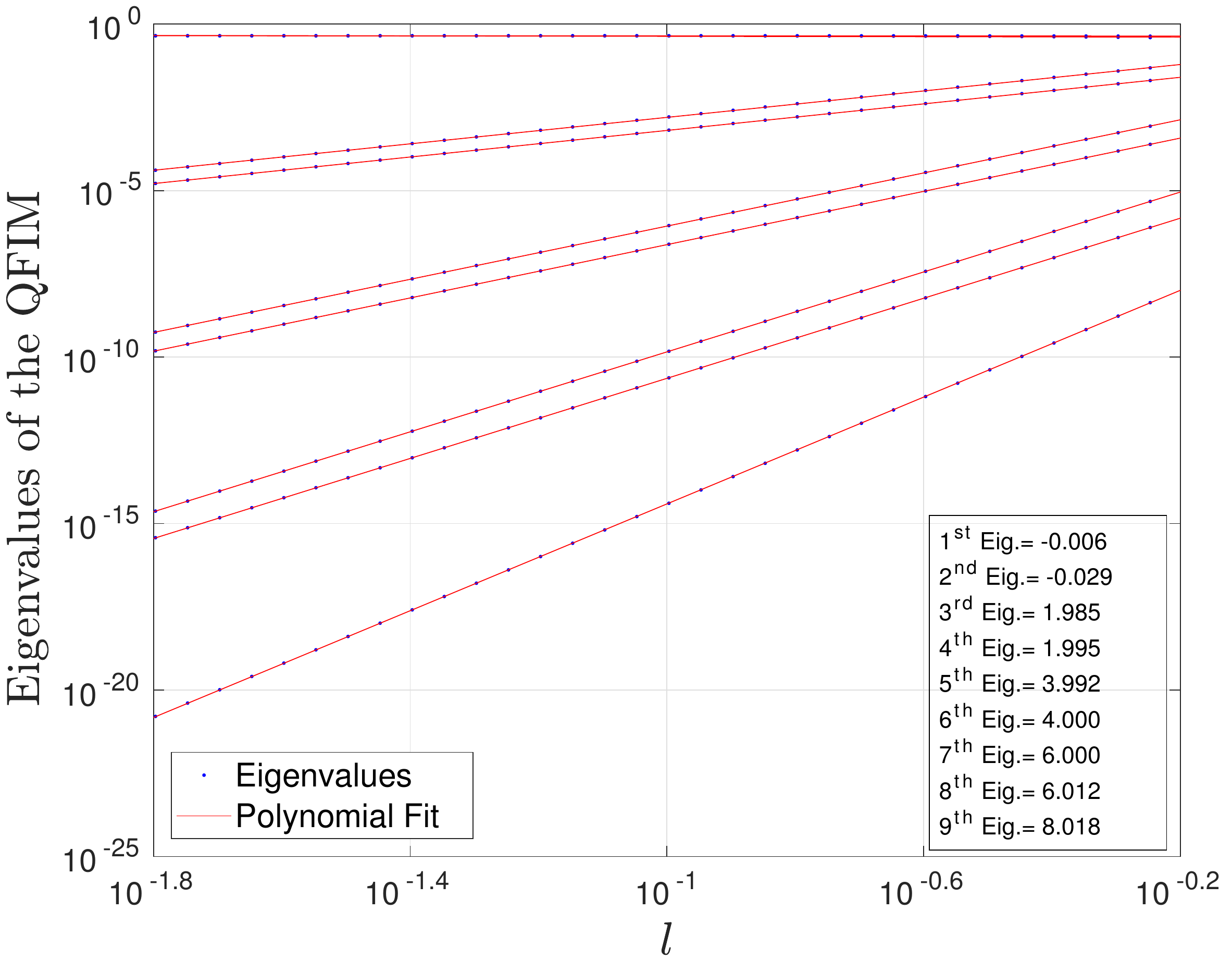}
\caption{\small Fitting of the eigenvalues of the QFIM matrix for the case of 9 sources in the limit of small distribution size. The sources are positioned at \( \alpha_i = i x \). The size of the distribution is denoted $l=8x$. The scale on both axes is logarithmic. The sources are separating from each other at equal distances, as in the previous plots. The slope of each line corresponding to different eigenvalues appears in the box in the plot.}
\label{fig:FITTING} 
\end{minipage}
~
\begin{minipage}[b]{0.49\textwidth}
\includegraphics[scale=0.38]{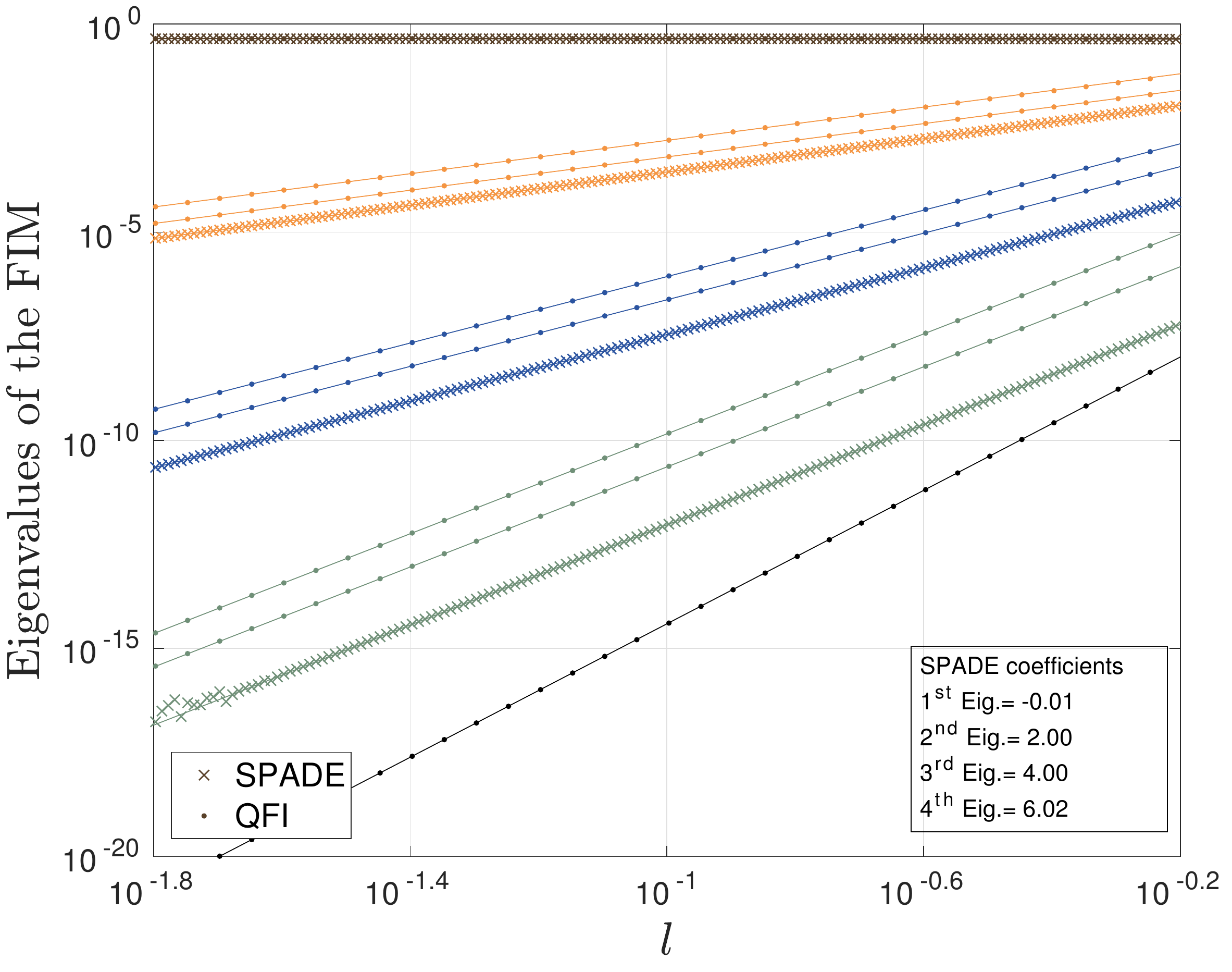}
	\caption{\small  The eigenvalues of the QFIM and CFIM for the SPADE with 20 HG modes and 9 equally bright sources. The sources are positioned at \( \alpha_i = (i-5)x \) such that the peak of \( \ket{\phi_0} \) is at the centroid of the distribution. The $x$ axis is the size $l$ of the distribution, with $l=8x$. The QFI eigenvalues scale as in Fig.~\ref{fig:FITTING}. By SPADE with 20 modes, we mean the POVM $ \{ \ket{\phi_{0}}\bra{\phi_{0}},\ket{\phi_{1}}\bra{\phi_{1}},\dots,\ket{\phi_{20}}\bra{\phi_{20}}, \mathbb{I} - \sum_{i=0}^{20}\ket{\phi_{i}} \bra{\phi_i} \}$. }
	\label{fig:CFIM} 
\end{minipage}
\end{figure}

Our analytical expression for the QFIM for localisation microscopy has enabled us to show that as point sources get closer, no more than two independent parameters can be estimated.
A rank-deficient QFIM occurs when the quantum state does not contain enough information to permit the estimation some of the parameters or combinations thereof.
The parameters that can be estimated correspond to the non-zero eigenvalues of the QFIM.
Without additional knowledge of the source distribution this restricts us to estimating functions of the first two moments \( f(\mathcal{C}_1,\mathcal{C}_2) \) only deep in the sub-Rayleigh limit. As Eq.~\eqref{eq:QFIm_qubit} shows, when all \( \{ \alpha_j \} \) are unknown as in localisation microscopy, there is vanishing information about any single \( \alpha_i \) itself.
This is in contrast to the scalar QFI $[\mathcal{Q}(\rho_{\bm{\alpha}})]_{ii}$ for $\alpha_{i}$ which is non zero, but assumes that all the other $\{ \alpha_{j} \}$ are known.
The manner in which the eigenvalues of the QFIM tend to zero is of interest in the search for optimal detection systems for localisation microscopy.
Numerical fitting in Fig.~\ref{fig:FITTING} shows the vanishing eigenvalues of the QFIM approach zero polynomially.
The degree of the polynomial is given by $d=2\lfloor\frac{\mu -1}{2}\rfloor$, where $\mu$ is the order the eigenvalue when arranged in descending order and $\lfloor \cdot \rfloor$ is the floor function.
These scalings are now extracted from the elements of the full QFIM of the localisation parameters $\bm{\alpha}$ -- rather than from bounds on estimating the various moments independently as in previous works~\cite{Zhou17,Tsang18QL}.

Unlike the latter, we can now compare the absolute performance of detection systems for localisation microscopy relative to its quantum limit. Indeed, while Fig.~\ref{fig:CFIM} shows the \( 2n \)-th eigenvalue of the QFIM closely parallel to the \( n \)-th eigenvalue of the CFIM for SPADE~\cite{TsaNL16}, there is a large gap in the absolute terms. This could be due to the sub-optimality of SPADE for estimating the \( \lfloor N/2\rfloor \) parameters it is sensitive to\footnote{Conventional SPADE is not sensitive to all the parameters needed to describe the sources' distribution, only its even moments~\cite{Tsang18QL,Zhou17,tsang_resolving_2019}.}. Similar scalings were observed with detection using superpositions of the conventional SPADE basis~\cite{Tsa17_IOP,Tsa17SPADE,Zhou17} that are sensitive to the other half of the moments. 
For reference over a range of separations,  Fig.~\ref{fig:CFIMs_eig} in Appendix \ref{app:CFIM} shows the eigenvalues of the CFIM for SPADE as well as direct imaging. Note that for both, the CFIM tends towards a rank one matrix.

\begin{figure}[H] 
 \includegraphics[scale=0.33]{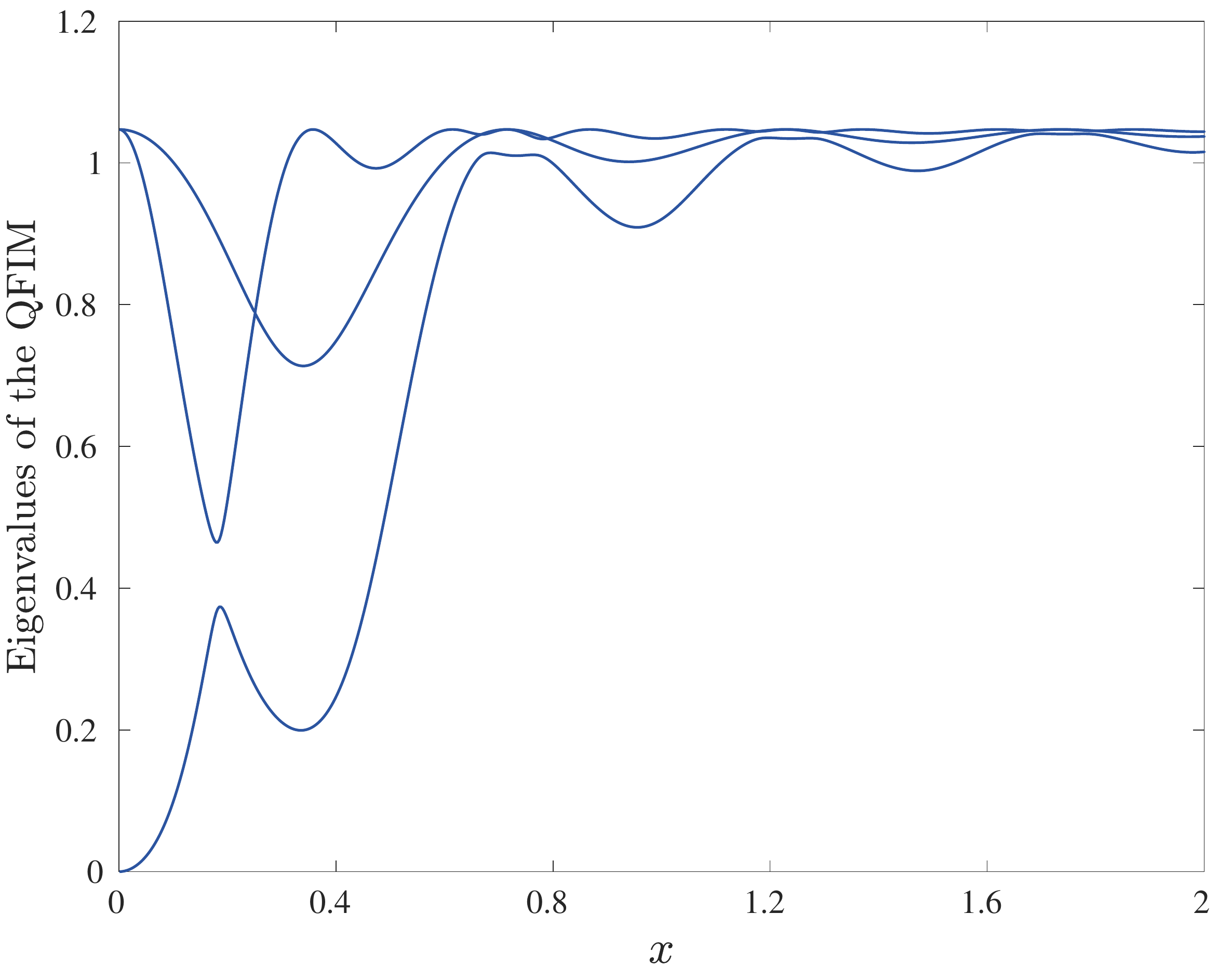}
\caption{\small The eigenvalues of the QFIM matrix in the case of 3 sources with equal intensities and a sinc PSF. The sources are separating from each other at equal distances, i.e.~$(\alpha_{1},\alpha_{2},\alpha_{3})=(x,2x,3x)$.}
\label{fig:sinc} 
\end{figure}

Finally, although our analytical result is derived with a Gaussian PSF, we expect the rank deficiency of the QFIM to be present in a more general family of PSFs.
To that end, Fig.~\ref{fig:sinc} shows the numerically obtained eigenvalues of the QFIM for three equidistant point sources of equal intensities under a sinc PSF (See Appendix~\ref{app:Sinc}) defined as
\begin{equation}
\psi_{\mathrm{PSF}}(x)=\frac{1}{\sqrt{\sigma}} \sinc \left( \frac{\pi x}{\sigma} \right)
\end{equation}
This PSF is the exact form for diffraction through a sharp one-dimensional slit which in its principal peak is well-approximated as Gaussian.

An approximation involving the first two spherical Bessel modes as in Sec.~\ref{sec:why_rank2} can be performed for a sinc PSF as well, leading to similar insights.
A proof of this rank deficiency for arbitrary PSFs and a physical explanation remains an open question.

To conclude, we have obtained several insights into the quantum limits of localisation microscopy via an analytical expression for the QFIM. In particular, the behaviour of the eigenvalues of the QFIM deep in the sub-Rayleigh limit revealed that only two parameters are eventually estimable. It also enabled us to compare the performance of known detection systems relative to the quantum limit in absolute terms, a question left open in the literature~\cite{tsang_resolving_2019}. The gap identified by us should motivate the search for detection systems, ideally on a single copy of the light field on the image plane, seeking to reduce or eliminate it. 

\section{Acknowledgements}
We thank Francesco Albarelli, Jamie Friel, Mankei Tsang, Liang Jiang, Lijian Zhang, Alex Retzker for enlightening discussions.
This study has been supported by the UK EPSRC (EP/K04057X/2),
the UK National Quantum Technologies Programme (EP/M01326X/1, EP/M013243/1) and the University of Warwick Global Partnership Fund.

\bibliography{Bibliography}

\setcounter{secnumdepth}{3}

\onecolumngrid

\appendix

\renewcommand{\theequation}{\Alph{section}.\arabic{equation}}
\counterwithin*{equation}{section}
\renewcommand{\thesubsection}{\thesection.\arabic{subsection}}%


\section{Expressing the density matrix in the HG basis} \label{app:HG_basis}
 
The density matrix is written in terms of the kets $\ket{\psi_{i}}$, which are expressed in the position space as in Eq.~\eqref{eq:rho_position}.
We assume a normalised Gaussian point spread function (PSF) of the form
 \begin{equation}\label{eq:psf}
	 \psi_{\PSF}(x)=\frac{1}{(2 \pi \sigma^{2})^{1/4}} e^{-\frac{x^{2}}{4\sigma^{2}}},
 \end{equation}
 and so
 \begin{equation}
	 \ket{\psi_i} = \int dx\, \psi_{\PSF}(x-\chi_i) \ket{x}.
 \end{equation}
The kets $\ket{\psi_{i}}$ can be expressed in terms of the complete Hermite-Gauss modes as
 \begin{equation} \label{eq:exp_HG}
	 \ket{\psi_{i}} = \sum_{q=0}^{\infty} \braket{\phi_{q}|\psi_{i}} \ket{\phi_{q}},
 \end{equation}
where $\ket{\phi_{q}}$ are the Hermite-Gauss modes, which can be expressed in the position space as~\cite{TsaNL16} 
 \begin{equation}\label{eq:HG_pos}
	 \ket{\phi_{q}}=\frac{1}{(2\pi \sigma^{2})^{1/4}} \frac{1}{\sqrt{2^{q}q!}} \int dx \mathrm{H}_{q} \left( \frac{x}{\sqrt{2} \sigma } \right) e^{- \frac{x^{2}}{4 \sigma^{2}} } \ket{x},
 \end{equation}
where $\mathrm{H}_{q}(x)$ are the Hermite polynomials.
The coefficients of the expansion Eq.~\eqref{eq:exp_HG} are 
 \begin{equation} \label{eq:APP_HGcoef}
 \begin{aligned}
 \braket{\phi_{q}|\psi_{i}}&=\frac{1}{\sqrt{2\pi \sigma^{2}}} \frac{1}{\sqrt{2^{q}q!}} \int dx dx' \mathrm{H}_{q} \left( \frac{x}{\sqrt{2} \sigma } \right) e^{- \frac{x^{2}}{4 \sigma} } e^{-\frac{(x'-\chi_{i})^{2}}{4\sigma^{2}}} \braket{x|x'}\\
 &=\frac{e^{-\frac{-\chi_{i}^{2}}{8\sigma^{2}}}}{\sqrt{2\pi \sigma^{2} 2^{q}q!}} \int dx\mathrm{H}_{q} \left( \frac{x}{\sqrt{2} \sigma } \right) e^{- (\frac{x}{\sqrt{2}\sigma} -\frac{\chi_{i}}{\sqrt{2} \sigma} )^{2} }\\
 &= \left( \frac{\chi_{i}}{2 \sigma}\right)^{q}\frac{e^{-\frac{1}{2}\left(\frac{\chi_{i}}{2 \sigma}\right)^2}}{\sqrt{q!}}
 \end{aligned}
 \end{equation}
Setting $\frac{\chi_{i}}{2 \sigma}= \alpha_{i}$ we get
 \begin{equation} \label{eq:coherentstate}
 \ket{\alpha_{i}} \equiv \ket{\psi_{i}}=\sum_{q=0}^{\infty} \frac{\alpha_{i}^{q}}{\sqrt{q!}} e^{-\alpha_{i}^{2}/2}  \ket{\phi_{q}}
 \end{equation}
 which has the same mathematical form as the coherent states with \( \{ \ket{\phi_q} \} \) forming the Fock basis~\cite{kok_lovett_2010}.

The state in Eq.~\eqref{eq:rho_position} can be also written in terms of the displacement operators $\mathcal{D}(\alpha_{i})=e^{\alpha_{i} (a^{\dagger}-a) }$, with $\alpha_{i} = \frac{\chi_{i}}{2 \sigma} \in \mathbb{R} $
 \begin{equation} \label{eq:rho_krauss}
 \rho_{\bm{\alpha}}=\sum_{i} \sqrt{w_{i}} \mathcal{D}(\alpha_{i}) \ket{0} \bra{0} \sqrt{w_{i}}\mathcal{D}^{\dagger}(\alpha_{i})
 \end{equation}
where $\mathcal{D}(\alpha)$ is the displacement operator.

The derivative of each coherent state with respect to its real amplitude \( \alpha \) is given by
\begin{equation}
\begin{aligned}
	\frac{\partial \ket{\alpha}}{\partial \alpha} & = \frac{\partial D(\alpha)}{\partial \alpha}  \ket{0}= \left(\hat{a}^{\dagger} -\alpha \right)\ket{\alpha}, \\
	\frac{\partial \bra{\alpha}}{\partial \alpha} &= \frac{\partial D^{\dagger}(\alpha)}{\partial \alpha}  \bra{0}= \left(\hat{a}-\alpha \right)\ket{\alpha},
\end{aligned}
\end{equation}
which yields the formula Eq.~\eqref{eq:rho_der}.

\section{Analytic results for \texorpdfstring{$N$}{N} sources} \label{app:Analyt}
The Tracy-Singh product~\cite{tracy_new_1972,KoNeWa91} defined for matrices \( A \) and \( B \) subdivided into blocks \( A_{ij} \) and \( B_{kl} \)
is $ A \TracySingh B $ where the \( (i,j) \)-th block of \( A \TracySingh B \) is \( A_{ij} \TracySingh B \) whose \( (k,l) \)-th block is in turn \( A_{ij} \otimes B_{kl} \).
That is if $A,B$ are block matrices with \begin{equation*}
 A = \begin{pmatrix}
A_{11} & A_{12} \\
A_{21} & A_{22}
\end{pmatrix},
\text{ and }
B = \begin{pmatrix}
B_{11} & B_{12} \\
B_{21} & B_{22}
\end{pmatrix},
\end{equation*}
then the Tracy-Singh product is
\begin{equation*}
A \odot B = \begin{pmatrix}[c|c]
A_{11} \odot B & A_{12} \odot B \\
\hline
A_{21} \odot B & A_{22} \odot B
\end{pmatrix} 
=
\begin{pmatrix}[cc|cc]
A_{11} \otimes B_{11} & A_{11} \otimes B_{12} & A_{12} \otimes B_{11} & A_{12} \otimes B_{12}\\
A_{11} \otimes B_{21} & A_{11} \otimes B_{22} & A_{12} \otimes B_{21} & A_{12} \otimes B_{22}\\
\hline
A_{21} \otimes B_{11} & A_{21} \otimes B_{12} & A_{22} \otimes B_{11} & A_{22} \otimes B_{12}\\
A_{21} \otimes B_{21} & A_{21} \otimes B_{22} & A_{22} \otimes B_{21} & A_{22} \otimes B_{22}
\end{pmatrix} 
\end{equation*}

Using the above definition, the matrix of Eq.~\eqref{eq:Lyap_gamma_fin} is found to be
\begin{equation} \label{eq:app_bba}
	\left( \Upsilon^{-1} \TracySingh \rho_{A} + \rho_{A} \TracySingh \Upsilon^{-1} \right)=
	\begin{bmatrix}
		D_{\bm{w}} \otimes \upsilon_{\alpha\alpha}+\upsilon_{\alpha\alpha} \otimes D_{\bm{w}} & D_{\bm{w}} \otimes \upsilon_{\alpha d} & \upsilon_{\alpha d} \otimes D_{\bm{w}} & 0 \\
		 D_{\bm{w}} \otimes \upsilon_{d\alpha} & D_{\bm{w}} \otimes \upsilon_{dd} & 0 &0 \\
		\upsilon_{d\alpha} \otimes D_{\bm{w}} & 0 & \upsilon_{dd} \otimes D_{\bm{w}}  &0 \\
		0& 0& 0& 0
\end{bmatrix}.
\end{equation}
where the elements of $\Upsilon^{-1}$ can be found using the formula of blockwise inversion:
\begin{equation}
\Upsilon^{-1}=\begin{bmatrix}
\upsilon_{\alpha\alpha} & \upsilon_{\alpha d} \\
\upsilon_{d\alpha} & \upsilon_{dd}
\end{bmatrix}
\end{equation}
with
\begin{equation} \label{eq:Y_in}
\begin{aligned}
\upsilon_{\alpha\alpha}&= \Upsilon_{\alpha\alpha}^{-1}+\Upsilon_{\alpha\alpha}^{-1}D_{\bm{\alpha}}\Upsilon_{\alpha\alpha} \left( \Upsilon_{dd} -\Upsilon_{\alpha\alpha}D_{\bm{\alpha}} \Upsilon_{\alpha\alpha}^{-1}D_{\bm{\alpha}} \Upsilon_{\alpha\alpha} \right)^{-1} \Upsilon_{\alpha\alpha}D_{\bm{\alpha}} \Upsilon^{-1} \\
\upsilon_{\alpha d}&=-\Upsilon_{\alpha\alpha}^{-1}D_{\bm{\alpha}} \Upsilon_{\alpha\alpha}  \left( \Upsilon_{dd} -\Upsilon_{\alpha\alpha}D_{\bm{\alpha}} \Upsilon_{\alpha\alpha}^{-1}D_{\bm{\alpha}} \Upsilon_{\alpha\alpha} \right)^{-1}\\
\upsilon_{d\alpha}&=-\left( \Upsilon_{dd} -\Upsilon_{\alpha\alpha}D_{\bm{\alpha}} \Upsilon_{\alpha\alpha}^{-1}D_{\bm{\alpha}} \Upsilon_{\alpha\alpha} \right)^{-1} \Upsilon_{\alpha\alpha}D_{\bm{\alpha}} \Upsilon_{\alpha\alpha} ^{-1}\\
\upsilon_{dd}&=\left( \Upsilon_{dd} -\Upsilon_{\alpha\alpha}D_{\bm{\alpha}} \Upsilon_{\alpha\alpha}^{-1}D_{\bm{\alpha}} \Upsilon_{\alpha\alpha} \right)^{-1}
\end{aligned}
\end{equation}
The inverse of the block matrix $\Upsilon_{\alpha \alpha}$ exists, because it is the Gramian matrix of the linear independent vectors $\ket{\alpha_{i}}$. 

For the QFIM elements we need to evaluate the inverse of the top left \( 3N^2 \times 3N^2 \) part of the matrix of Eq.~\eqref{eq:app_bba} which we denote \( \mathbb{A} \).
In order to obtain the inverse of $\mathbb{A}$, we need to further partition $\mathbb{A}$ as
\begin{equation}
\mathbb{A}=\begin{bmatrix}
\varepsilon & \vartheta\\
\varphi & \varpi
\end{bmatrix}
\end{equation}
with
\begin{equation}
\begin{aligned}
\varepsilon &= \begin{bmatrix} D_{\bm{w}} \otimes \upsilon_{\alpha\alpha}+\upsilon_{\alpha\alpha} \otimes D_{\bm{w}} \end{bmatrix} &
\vartheta &= \begin{bmatrix} D_{\bm{w}} \otimes \upsilon_{\alpha d} & \upsilon_{\alpha d} \otimes D_{\bm{w}} \end{bmatrix} \\
\varphi &= \begin{bmatrix}
D_{\bm{w}} \otimes \upsilon_{d\alpha}\\
\upsilon_{d\alpha} \otimes D_{\bm{w}} 
\end{bmatrix} &
\varpi &= \begin{bmatrix}
 D_{\bm{w}} \otimes \upsilon_{dd} & 0 \\
0 & \upsilon_{dd} \otimes D_{\bm{w}} 
\end{bmatrix}
\end{aligned}  
\end{equation}
The inverse of $\varpi$ is
\begin{equation}
	\varpi^{-1} = 
\begin{pmatrix}
 D_{\bm{w}}^{-1}\otimes \upsilon^{-1}_{dd} & 0 \\
 0 & \upsilon^{-1}_{dd}\otimes D_{\bm{w}}^{-1} \\
\end{pmatrix}
\end{equation}
The elements of $\mathbb{A}^{-1}$ will be given by the formulas
\begin{equation}
\begin{aligned}
(\mathbb{A}^{-1})_{11} &= \left(\varepsilon - \vartheta \varpi^{-1} \varphi \right)^{-1} = S^{-1} \\
(\mathbb{A}^{-1})_{12} &= - S^{-1} \vartheta \varpi^{-1}   \\
(\mathbb{A}^{-1})_{21} &=  - \varpi^{-1} \varphi S^{-1} \\
(\mathbb{A}^{-1})_{22} &= \varpi^{-1}+ \varpi^{-1} \varphi S^{-1} \vartheta \varpi^{-1}
\end{aligned}
\end{equation}
After calculations and by substituting the $\Upsilon^{-1}$ elements from Eq.~\eqref{eq:Y_in}, we derive the explicit form of $\mathbb{A}^{-1}$ elements:
\begin{equation} \label{eq:app_abb_inv}
\begin{aligned}
(\mathbb{A}^{-1})_{11} &= S^{-1} = \left( \Upsilon_{\alpha\alpha}^{-1} \otimes D_{\bm{w}} +D_{\bm{w}} \otimes \Upsilon_{\alpha\alpha}^{-1} \right)^{-1}\\ 
(\mathbb{A}^{-1})_{12} &= S^{-1} \begin{pmatrix}
\mathbb{I} \otimes (\Upsilon_{\alpha\alpha}^{-1} \Upsilon_{\alpha d}) & (\Upsilon_{\alpha\alpha}^{-1} \Upsilon_{\alpha d}) \otimes \mathbb{I}
\end{pmatrix}  \\
(\mathbb{A}^{-1})_{21} &= \begin{pmatrix}
\mathbb{I} \otimes (\Upsilon_{d\alpha} \Upsilon_{\alpha\alpha}^{-1})\\
 (\Upsilon_{d\alpha} \Upsilon_{\alpha\alpha}^{-1}) \otimes \mathbb{I}
\end{pmatrix} S^{-1} \\
(\mathbb{A}^{-1})_{22} &= \begin{pmatrix}
D_{\bm{w}}^{-1}\otimes \upsilon^{-1}_{dd} & 0 \\
 0 & \upsilon^{-1}_{dd}\otimes D_{\bm{w}}^{-1} \\
\end{pmatrix} 
+ \begin{pmatrix}
\mathbb{I} \otimes (\Upsilon_{d\alpha} \Upsilon_{\alpha\alpha}^{-1})\\
 (\Upsilon_{d\alpha} \Upsilon_{\alpha\alpha}^{-1}) \otimes \mathbb{I}
\end{pmatrix} S^{-1} \begin{pmatrix}
\mathbb{I} \otimes (\Upsilon_{\alpha\alpha}^{-1} \Upsilon_{\alpha d}) & (\Upsilon_{\alpha\alpha}^{-1} \Upsilon_{\alpha d}) \otimes \mathbb{I}
\end{pmatrix}
\end{aligned}
\end{equation}

The QFIM elements are then obtained from Eq.~\eqref{eq:gamma_with_inverse} and~\eqref{eq:app_abb_inv}
\begin{equation}
\begin{aligned}
	\mathcal{Q}_{ij} &= 2 w_{i} w_{j} \begin{bmatrix}
-2 \alpha_{i} \Vectr{E_{i}} & \Vectr{E_{i}} & \Vectr{E_{i}}
\end{bmatrix} 
 \mathbb{A}^{-1} \begin{bmatrix}
 -2 \alpha_{j} \Vect{E_{j}}\\
 \Vect{E_{j}} \\
 \Vect{E_{j}}
\end{bmatrix} \\
&= 2 w_{i} w_{j} \Vectr{E_{i}} \begin{bmatrix}
	-2\alpha_i \mathbb{I} \otimes \mathbb{I} & \mathbb{I} \otimes \mathbb{I} & \mathbb{I} \otimes \mathbb{I}
\end{bmatrix} 
 \mathbb{A}^{-1} \begin{bmatrix}
	 -2\alpha_j \mathbb{I} \otimes \mathbb{I} \\ \mathbb{I} \otimes \mathbb{I}\\ \mathbb{I} \otimes \mathbb{I}
\end{bmatrix} 
\Vect{E_{j}} \\
&= 2 w_{i} w_{j} 
\Vectr{E_{i}} \left[
\mathbb{I} \otimes \Upsilon_{d\alpha} \Upsilon_{\alpha\alpha}^{-1} + \Upsilon_{d\alpha} \Upsilon_{\alpha\alpha}^{-1} \otimes \mathbb{I} - 2\alpha_i \mathbb{I} \otimes \mathbb{I} \right] 
 S^{-1}
 \left[
\mathbb{I} \otimes \Upsilon_{\alpha\alpha}^{-1} \Upsilon_{\alpha d} + \Upsilon_{\alpha\alpha}^{-1} \Upsilon_{\alpha d} \otimes \mathbb{I} - 2\alpha_j \mathbb{I} \otimes \mathbb{I} \right] 
\Vect{E_j} \\
&\mkern16mu +
2 w_i w_j
\Vectr{E_i} E_{\bm{w}}^{-1} \otimes \upsilon^{-1}_{dd} + \upsilon^{-1}_{dd} \otimes E_{\bm{w}}^{-1} \Vect{E_j} \\
&= 2 w_{i} w_{j} 
\Vectr{E_{i}} \left[
\mathbb{I} \otimes \Upsilon_{d\alpha} \Upsilon_{\alpha\alpha}^{-1} + \Upsilon_{d\alpha} \Upsilon_{\alpha\alpha}^{-1} \otimes \mathbb{I} - 2\alpha_i \mathbb{I} \otimes \mathbb{I} \right] 
 S^{-1}
 \left[
\mathbb{I} \otimes \Upsilon_{\alpha\alpha}^{-1} \Upsilon_{\alpha d} + \Upsilon_{\alpha\alpha}^{-1} \Upsilon_{\alpha d} \otimes \mathbb{I} - 2\alpha_j \mathbb{I} \otimes \mathbb{I} \right] 
\Vect{E_j} \\
&\mkern16mu +
4 w_i \delta_{ij}
(\upsilon^{-1}_{dd})_{ij}\\
&= 2 w_{i} w_{j} 
\Vectr{E_{i}} \left[
\mathbb{I} \otimes \Upsilon_{d\alpha} \Upsilon_{\alpha\alpha}^{-1} + \Upsilon_{d\alpha} \Upsilon_{\alpha\alpha}^{-1} \otimes \mathbb{I} - 2\alpha_i \mathbb{I} \otimes \mathbb{I}
\right] S^{-1} \left[
\mathbb{I} \otimes \Upsilon_{\alpha\alpha}^{-1} \Upsilon_{\alpha d} + \Upsilon_{\alpha\alpha}^{-1} \Upsilon_{\alpha d} \otimes \mathbb{I} - 2\alpha_j \mathbb{I} \otimes \mathbb{I}
\right] \Vect{E_j} \\
&\mkern16mu +
4 w_i \delta_{ij}
\left[ 1+\alpha_i^2 - (\Upsilon_{\alpha\alpha}D_{\alpha}\Upsilon_{\alpha\alpha}^{-1}D_{\alpha}\Upsilon_{\alpha\alpha})_{ij} \right]
\end{aligned}
\end{equation}

\begin{figure} [htbp]
\centering
 \subfloat[]{
 \includegraphics[scale=0.38]{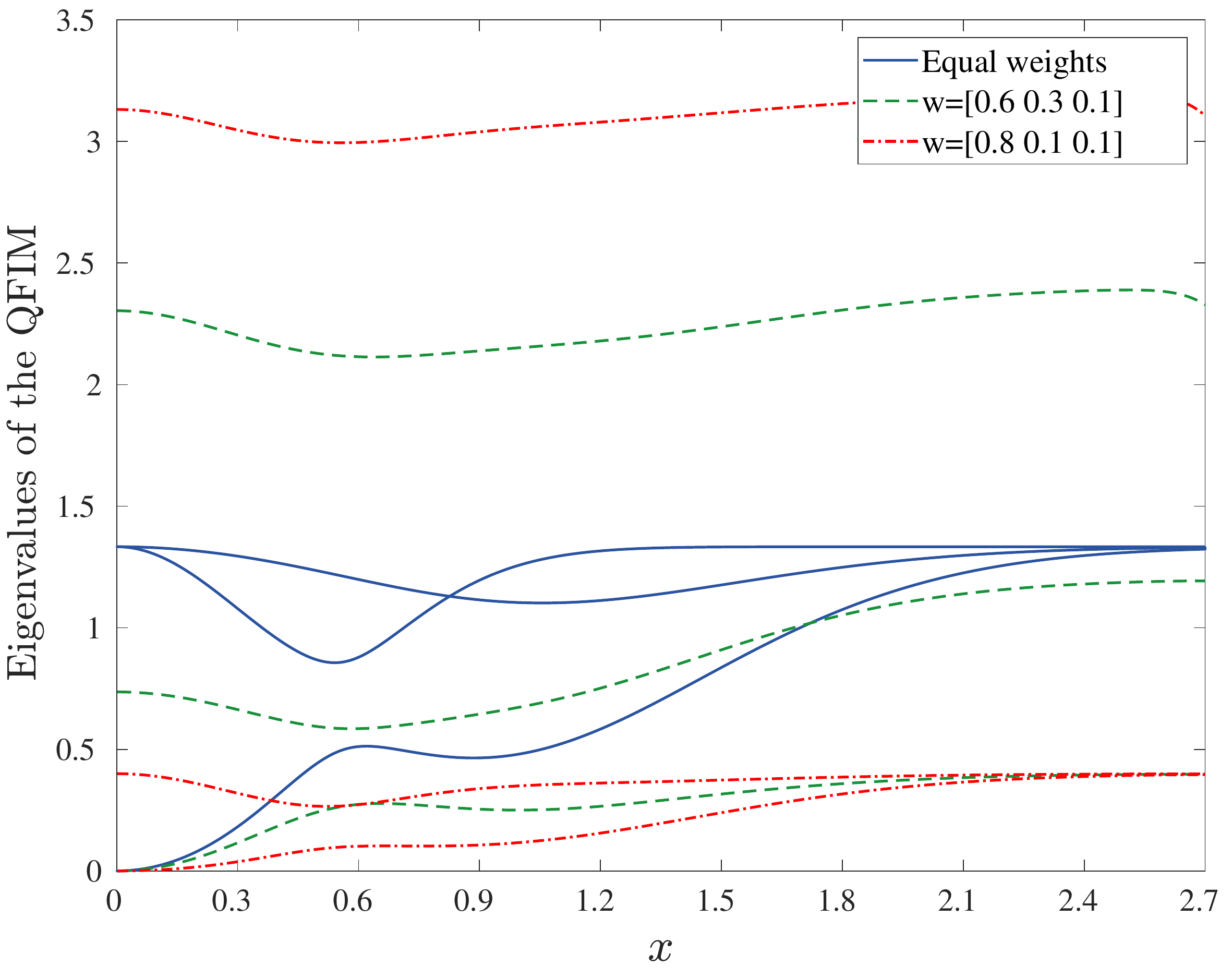}
 }
 \subfloat[]{ \includegraphics[scale=0.38]{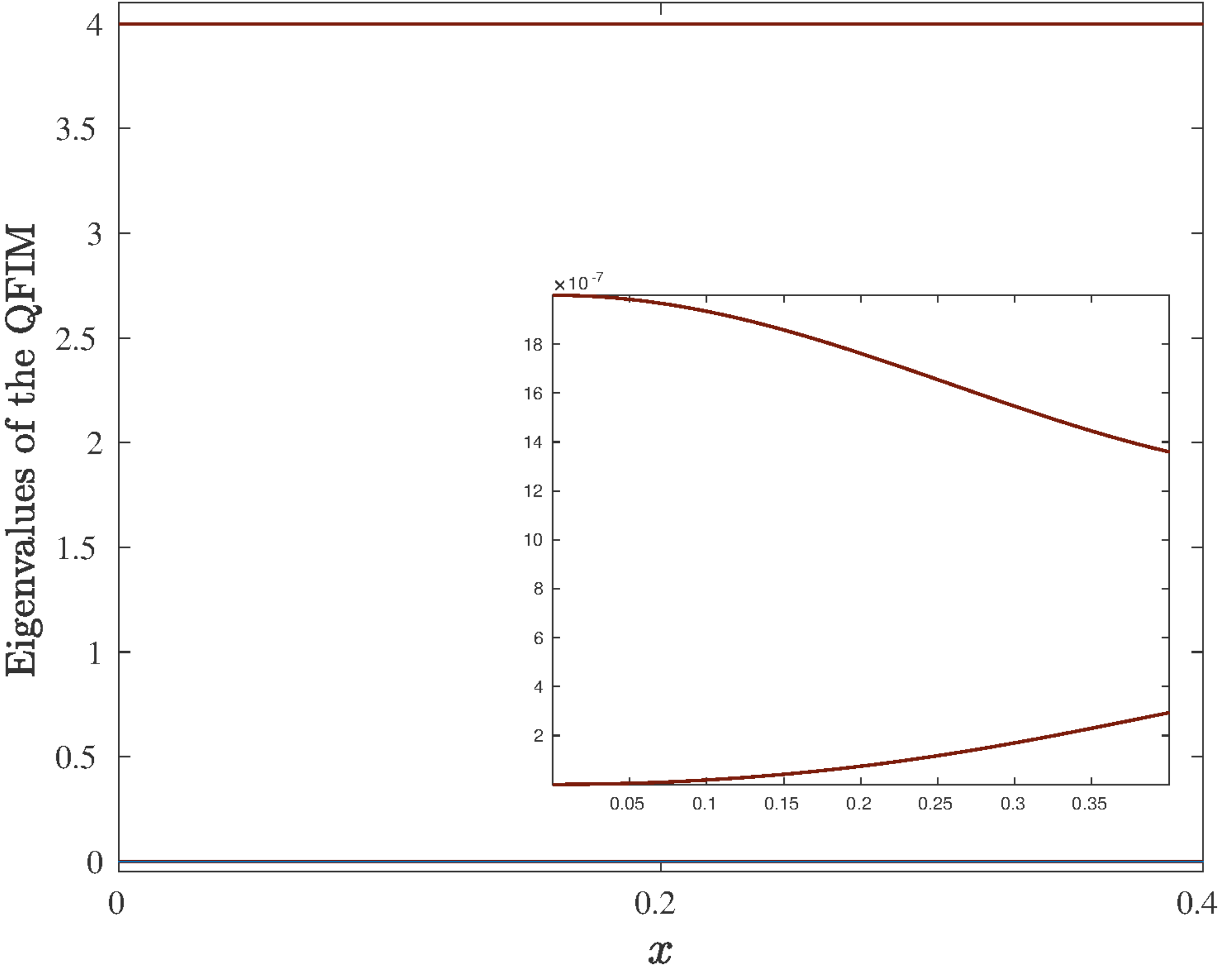}}
\caption{\small The eigenvalues of the QFIM matrix in the case of 3 sources. The sources are separating from each other at equal distances, i.e.~$(\alpha_{1},\alpha_{2},\alpha_{3})=(x,2x,3x).$ It can be noticed that the limiting values of the two non zero eigenvalues are different as the weights become different. However, the rank 2 of the QFIM remains. In Fig.~(b) the limiting case of one extremely bright source wand two very weak ones is displayed. The inset shows the two vanishing eigenvalues. }
 \label{fig:qfi_unequal_weights}
 \end{figure}

Finally, to complement the discussion in the main text,we  present some further examples of the QFIM eigenvalues for $N=4,5$ spurces and in Fig.~\ref{fig:qfi_unequal_weights} we present the eigenvalues of the QFIM for 3 sources in the case of unequal weights (relative intensities)~Fig.(\ref{fig:qfi_unequal_weights}). 

\begin{figure} [htbp]
	\centering

\subfloat{
\includegraphics[scale=0.38]{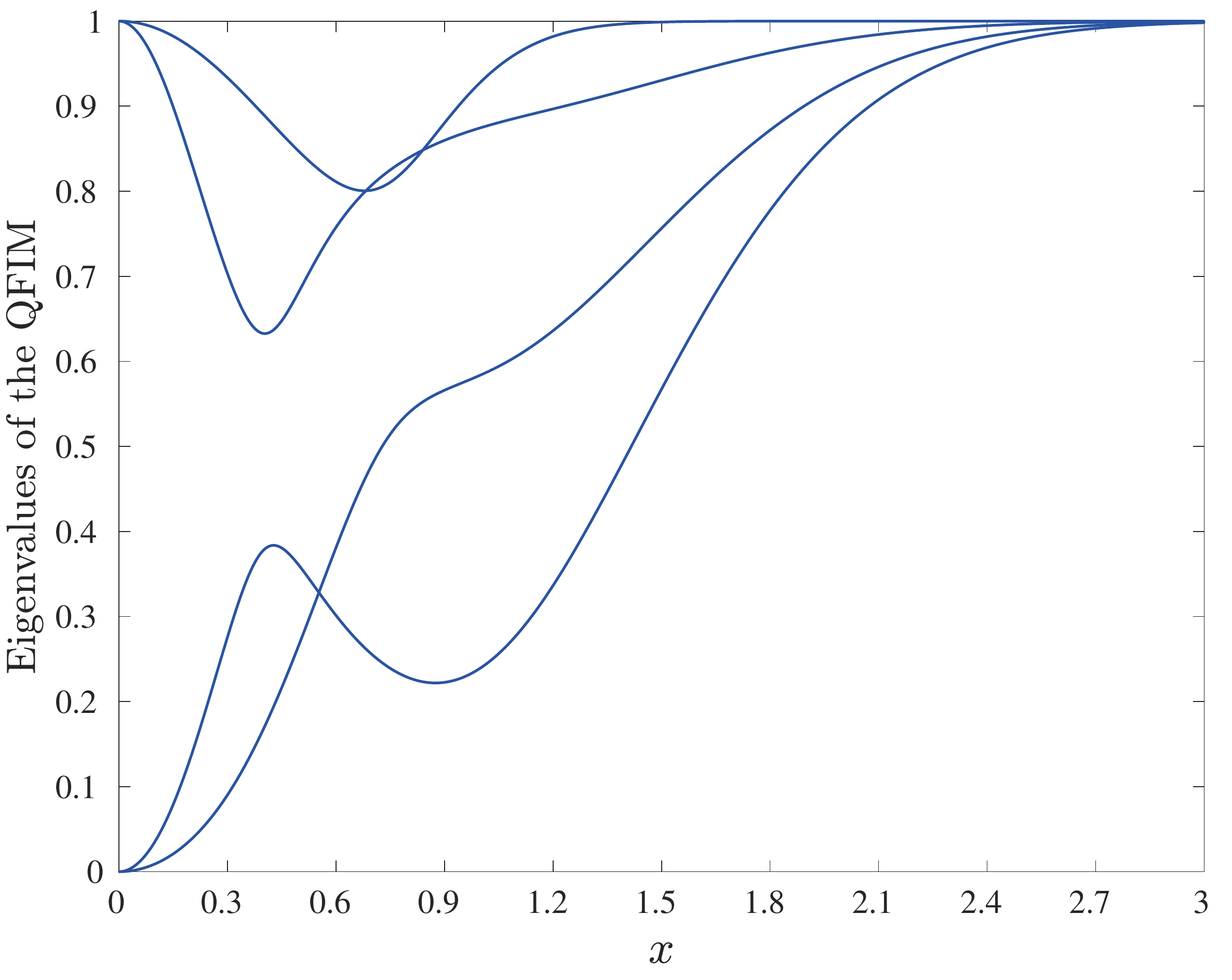}
}
\subfloat{
\includegraphics[scale=0.38]{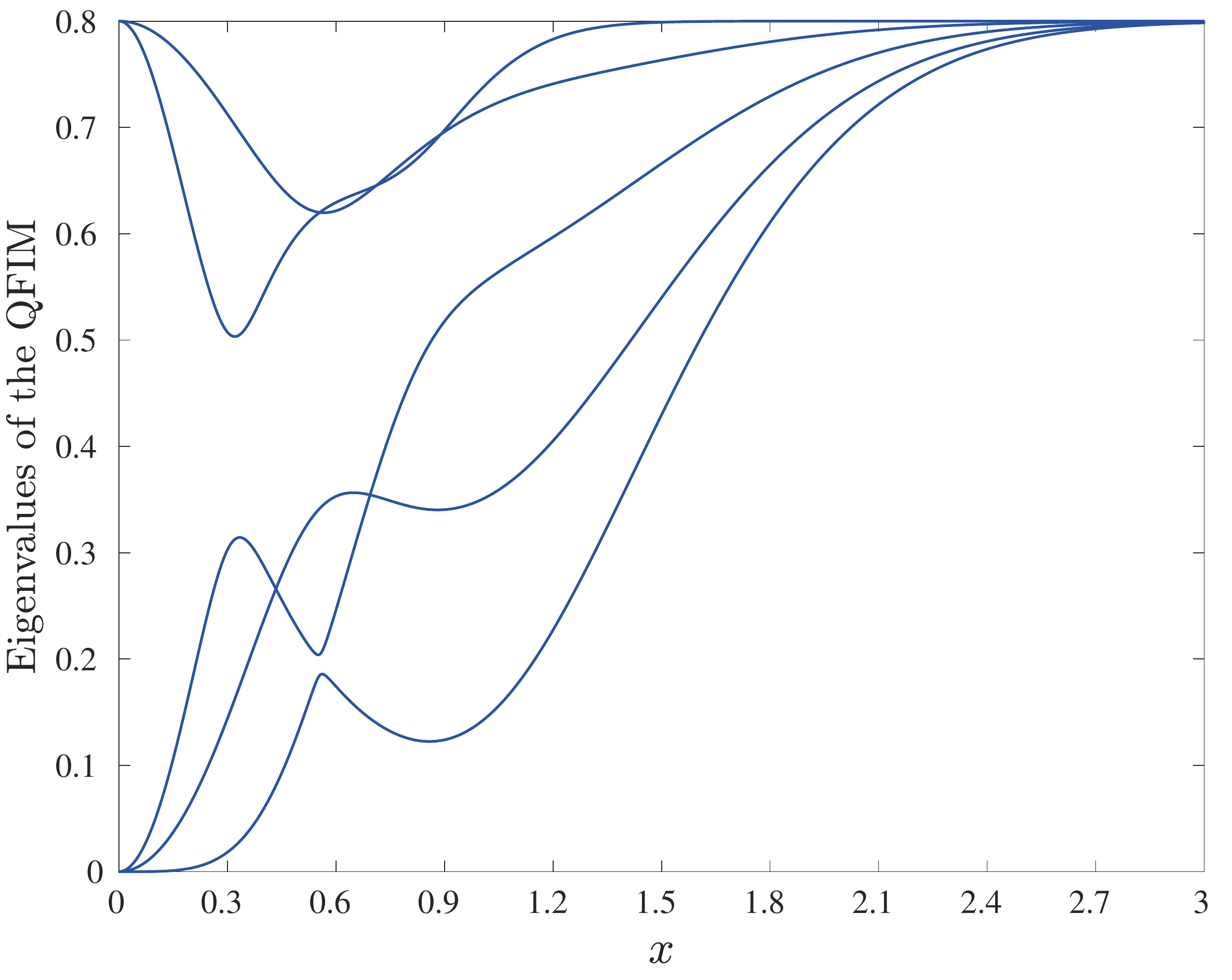}
}
\caption{\small The eigenvalues of the QFIM for 4 (left) and 5 (right) sources with equal intensities. The sources are separating from each other by equal distances: $(\alpha_{1},\alpha_{2},\alpha_{3},\alpha_{4})=(x,2x,3x,4x)$ and $(\alpha_{1},\alpha_{2},\alpha_{3},\alpha_{4},\alpha_{5})=(x,2x,3x,4x,5x)$.}  \label{fig:eig45}
\end{figure}

\section{Analytic results for \texorpdfstring{$x_{i} \ll \sigma$}{x<<σ}} \label{app:eigen}

The state in the sub-diffraction regime is given by Eq.~\eqref{eq:rho_qubit}. The derivative can be calculated immediately from this formula and it is
\begin{equation}
\frac{\partial}{\partial \alpha_{i}} \rho =-2\alpha_{i} \ket{0}\bra{0}+ \left[ \ket{0}\bra{1}+\ket{1}\bra{0} \right] + 2\alpha_{i} \ket{1} \bra{1}=\begin{bmatrix}
-2\alpha_{i} & 1\\
1 & 2\alpha_{i}
\end{bmatrix}
\end{equation}
By solving the SLD equation $\partial_{a_i} \rho_{\bm{\alpha}} = (\rho_{\bm{\alpha}} L_{i}+L_{i} \rho_{\bm{\alpha}})$, we can determine the SLDs in the $\{\ket{0},\ket{1} \}$ basis:
\begin{equation}
L_{i}=\frac{2}{(\mathcal{C}_{2} -1)\mathcal{C}_{2}+\mathcal{C}_{1}^{2}} \begin{bmatrix}
\mathcal{C}_{2} \mathcal{C}_{1} +(\mathcal{C}_{2}-\mathcal{C}_{1}^{2})\alpha_{i} & (\mathcal{C}_{2} -1)\mathcal{C}_{2}+(\mathcal{C}_{1}-2\mathcal{C}_{2}\mathcal{C}_{1})\alpha_{i}\\
(\mathcal{C}_{2} -1)\mathcal{C}_{2}+(\mathcal{C}_{1}-2\mathcal{C}_{2}\mathcal{C}_{1})\alpha_{i} & \mathcal{C}_1-\mathcal{C}_{1} \mathcal{C}_{2} +(2\mathcal{C}_{1}^{2}+\mathcal{C}_{2}-1)\alpha_{i}
\end{bmatrix}
\end{equation}
Knowing the SLDs, we can obtain the QFIM of Eq.~\eqref{eq:QFIm_qubit}.

As already mentioned in the main text, the rank of the QFIM only depends on the matrix 
\begin{equation}\label{eq:pinakas}
\begin{bmatrix}M_{11} & M_{12} \\ M_{21} & M_{22}
 \end{bmatrix}
\end{equation}
of Eq.~\eqref{eq:QFIm_qubit}, with the elements of this matrix given by Eq.~\eqref{eq:qfi_qubit_coef}. 
The eigenvalues $\mu_{1}, \mu_{2}$ of the matrix Eq.~\eqref{eq:pinakas} are
\begin{equation}
\begin{aligned}
\mu_{1} &= \frac{1}{2} \left(\mathcal{C}_{2} ^{2}-\sqrt{\left((\mathcal{C}_{2} -1) \mathcal{C}_{2} +4 \mathcal{C}_{1} ^{2}-1\right)^{2}+4 \left((\mathcal{C}_{2} -1) \mathcal{C}_{2} +\mathcal{C}_{1} ^{2}\right)}-\mathcal{C}_{2} +4 \mathcal{C}_{1} ^{2}-1\right),\\
\mu_{2} &= \frac{1}{2} \left(\mathcal{C}_{2} ^{2}+\sqrt{\left((\mathcal{C}_{2} -1) \mathcal{C}_{2} +4 \mathcal{C}_{1} ^{2}-1\right)^{2}+4 \left((\mathcal{C}_{2} -1) \mathcal{C}_{2} +\mathcal{C}_{1} ^{2} \right)}-\mathcal{C}_{2} +4 \mathcal{C}_{1} ^{2}-1\right)
\end{aligned}
\end{equation}
The condition for the eigenvalues to be zero is
\begin{equation} \label{eq:log_cond}
\left(0=3 \mathcal{C}_{1} ^2-1-\sqrt{\left(3 \mathcal{C}_{1} ^2-1\right)^2}\land \mathcal{C}_{2} =\frac{1}{2} \left(1-\sqrt{1-4 \mathcal{C}_{1} ^2}\right)\right)\lor \left(0=3 \mathcal{C}_{1} ^2-1-\sqrt{\left(3 \mathcal{C}_{1} ^2-1\right)^2}\land \mathcal{C}_{2} =\frac{1}{2} \left(1+\sqrt{1-4 \mathcal{C}_{1} ^2}\right)\right)
\end{equation}
The first part $0=3 \mathcal{C}_{1} ^2-1-\sqrt{\left(3 \mathcal{C}_{1} ^2-1\right)^2}$ is always true, as it reduces to the identity $\left(3 \mathcal{C}_{1} ^2-1\right)^2=\left(3 \mathcal{C}_{1} ^2-1\right)^2$. For the second part we have
\begin{equation}
\begin{aligned}
\mathcal{C}_{2} =\frac{1}{2} \left(1\pm \sqrt{1-4 \mathcal{C}_{1} ^2}\right) \Leftrightarrow \left( 2\mathcal{C}_{2} -1 \right)^{2}=1-4\mathcal{C}_{1}^{2} \Leftrightarrow \mathcal{C}_{2}^{2}-\mathcal{C}_{2} +\mathcal{C}_{1}^{2} =0
\end{aligned}
\end{equation}
Substituting $\mathcal{C}_{2}$ and $\mathcal{C}_{1}$ we get
\begin{equation}
\begin{aligned}
& \left( \sum_{i=1}^{N} \alpha_{i}^{2} \right)^{2} -\sum_{i=1}^{N} \alpha_{i}^{2} +\left(\sum_{i=1}^{N} \alpha_{i} \right) ^{2} =0 \Leftrightarrow 
\left( \sum_{i=1}^{N} \alpha_{i}^{2} \right)^{2} -\sum_{i=1}^{N} \alpha_{i}^{2}+\sum_{i=1}^{N} \alpha_{i}^{2}  +2\sum_{i,j=1,i\neq j}^{N} \alpha_{i}\alpha_{j} =0 \\
& \left( \sum_{i=1}^{N} \alpha_{i}^{2} \right)^{2} +2\sum_{i,j=1,i\neq j}^{N} \alpha_{i}\alpha_{j} =0
\end{aligned}
\end{equation}
Since $\alpha_{i}$ are strictly positive, except one that can be zero, this sum of positive terms cannot be equal to zero. Therefore, this statement is always false.
Thus, the Eq.~\eqref{eq:log_cond} becomes $(1 \land 0) \lor (1 \land 0)=0$, which means that the two eigenvalues can never be zero and the QFIM will be rank 2.


\section{Calculation of the QFI for the Sinc PSF} \label{app:Sinc}
 
The expansion of the Sinc function on the HG modes is not ideal for numerical calculations.
Instead we use the spherical Bessel function of the $1^{\mathrm{st}}$ kind and express the states onto those modes in which we then truncate.
If the PSF is a $\sinc$ function, the $\ket{\psi_{i}}$ are
\begin{equation}
\ket{\psi_{i}}=\frac{1}{\sqrt{\sigma}}\int_{-\infty}^{\infty} \sinc \left( \frac{\pi (x-X_{i})}{\sigma} \right) \ket{x}
\end{equation}
We can use the identity~\cite{AbSt} 
\begin{equation}\label{eq:app_sincid}
\sinc\left( \frac{\pi (x-x')}{\sigma} \right) =\sum_{q=0}^{\infty} (2q+1) J_{q}\left( \frac{\pi x}{\sigma} \right) J_{q}\left( \frac{\pi x'}{\sigma} \right),
\end{equation}
where $J_{q}(x)$ is the spherical Bessel function of the $1^{st}$ kind. The spherical Bessel function are orthogonal in all $\mathbb{R}$
\begin{equation}
\int_{-\infty}^{\infty} dx \, J_{q}(x) J_{p}(x) = \frac{\pi}{2q+1} \delta_{qp},
\end{equation}
therefore we can define the orthonormal basis
\begin{equation}
\ket{j_{q}}= \sqrt{ \frac{2q+1}{\sigma}} \int_{-\infty}^{\infty} dx \, J_{q} \left( \frac{\pi x}{\sigma} \right) \ket{x}
\end{equation}
The set of the spherical Bessel functions is a basis in $\mathbb{R}$, but is not complete since it is not a resolution of identity as we can see from Eq.~\eqref{eq:app_sincid}.
Hence, we can expand the sinc function on the bessel function basis, using the identity Eq.~\eqref{eq:app_sincid}:
\begin{equation}
\begin{aligned}
\ket{\psi_{i}}&=\frac{1}{\sqrt{\sigma}}\int_{-\infty}^{\infty}\sum_{q=0}^{\infty} (2q+1) J_{q}\left( \frac{\pi x}{\sigma} \right) J_{q}\left( \frac{\pi X_{i}}{\sigma} \right) \ket{x} \\
&=\sum_{q=0}^{\infty} \sqrt{2q+1} \, J_{q} \left( \frac{\pi X_{i}}{\sigma} \right)  \ket{j_{q}}
\end{aligned}
\end{equation}
Using the identity for the Bessel functions
\begin{equation}
\frac{\partial J_{q}(x)}{\partial x}=J_{q-1}(x)-\frac{q+1}{2}J_{q}(x)
\end{equation}
we can also have an expression for the derivative of $\ket{\psi_{i}}$
\begin{equation}
\frac{\partial \ket{\psi_{i}}}{\partial X_{i}}=\frac{\pi}{\sigma} \left( J_{q-1} \left( \frac{\pi X_{i}}{\sigma} \right)-\frac{q+1}{2}J_{q} \left( \frac{\pi X_{i}}{\sigma} \right) \right)
\end{equation}
We see that both the state $\rho$ and its derivatives are completely expressed within the basis $\ket{j_{q}}$.
This means that we can use the definition of the SLD (Eq.~\ref{eq:SLd_orismos}) and express the SLD in the same basis.
\begin{equation} \label{eq:SLd_orismos}
2 \frac{\partial \rho}{\partial \alpha_{\mu}}= \rho L^{\mu} + L^{\mu} \rho
\end{equation}
In this way the fact that the specific basis is not complete does not affect our calculations.

\begin{figure}[H]
\centering
\includegraphics[scale=0.4]{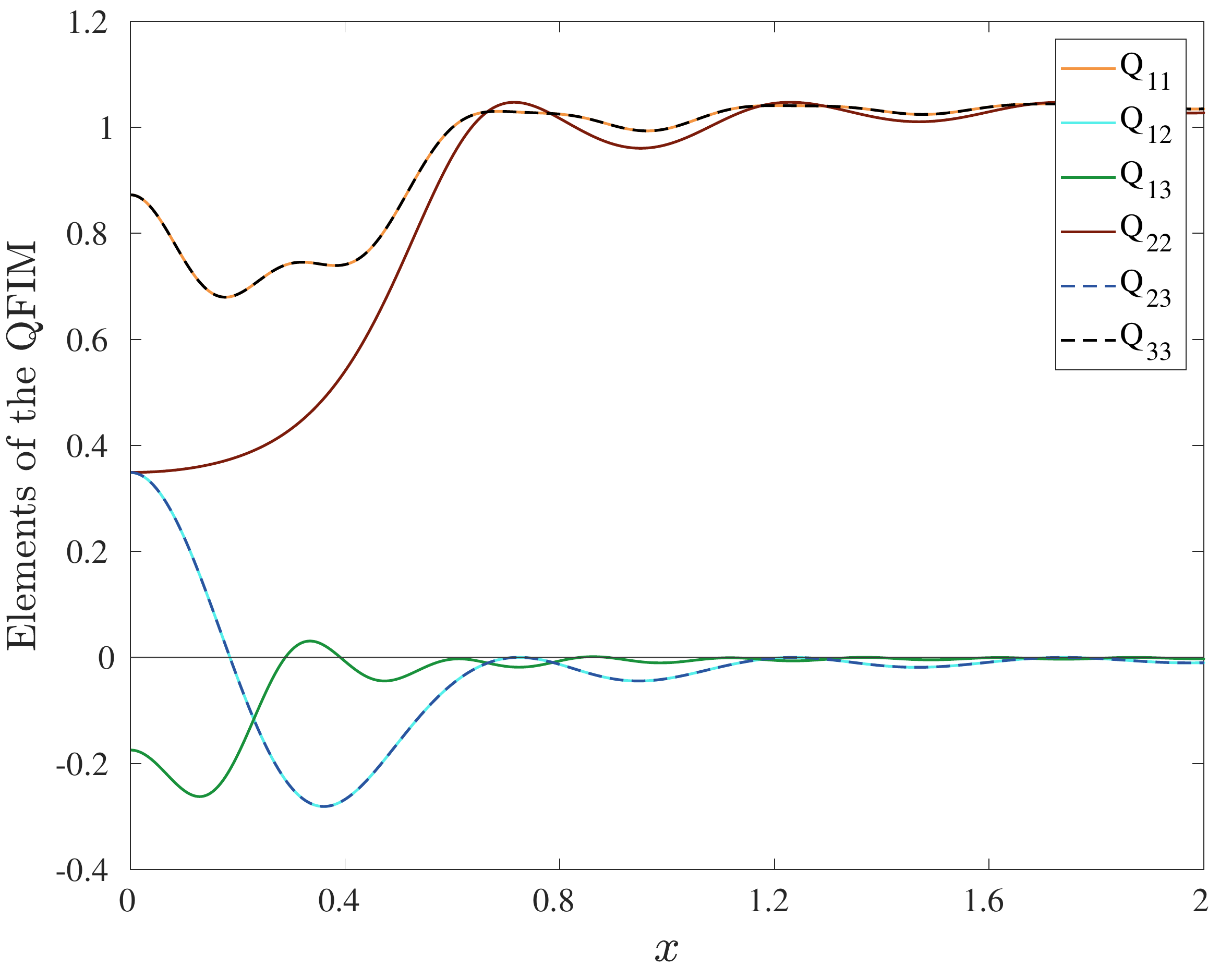}
\caption{\small The eigenvalues of the QFIM matrix in the case of 3 sources with equal intensities for the sinc PSF. The sources are separating from each other at equal distances, i.e.~$(\alpha_{1},\alpha_{2},\alpha_{3})=(x,2x,3x)$.}
\label{fig:sinc_elem}
\end{figure}

For the numerical calculations we have to truncate our state in the appropriate amount of modes. From Figs.~\ref{fig:sinc} and~\ref{fig:sinc_elem}, we can see that our conclusions do not change with the use of a non-Gaussian PSF. 

\section{Eigenvalues of the CFIM for SPADE and Direct Imaging} \label{app:CFIM}

Finally, we present the eigenvalues of the CFIM for SPADE and direct imaging fir a large range of separations. 

\begin{figure} [htbh]
	\subfloat[
 \label{fig:SP_eig}]{
 \includegraphics[scale=0.38]{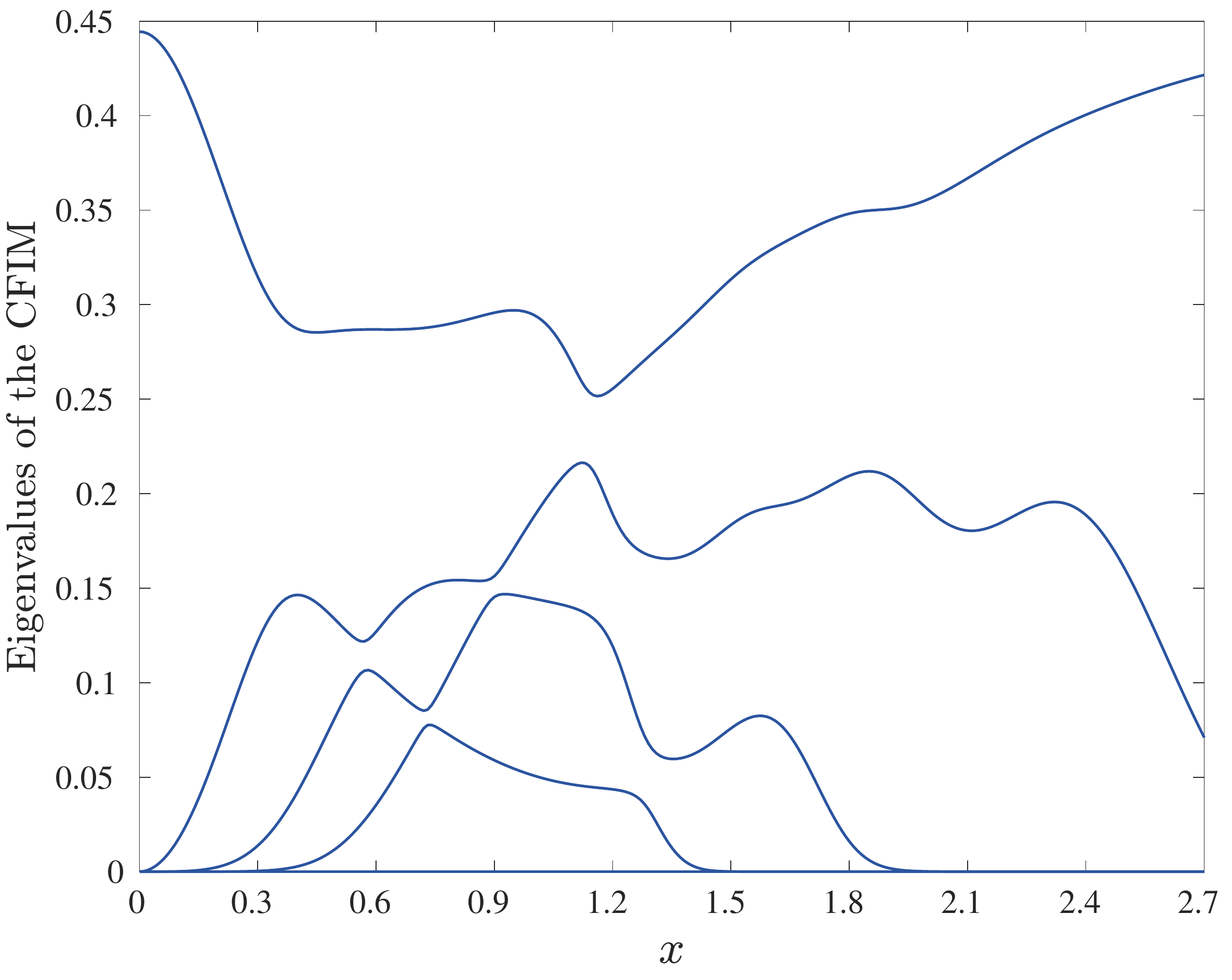}
 }
\subfloat[
\label{fig:DI_eig}]{
 \includegraphics[scale=0.38]{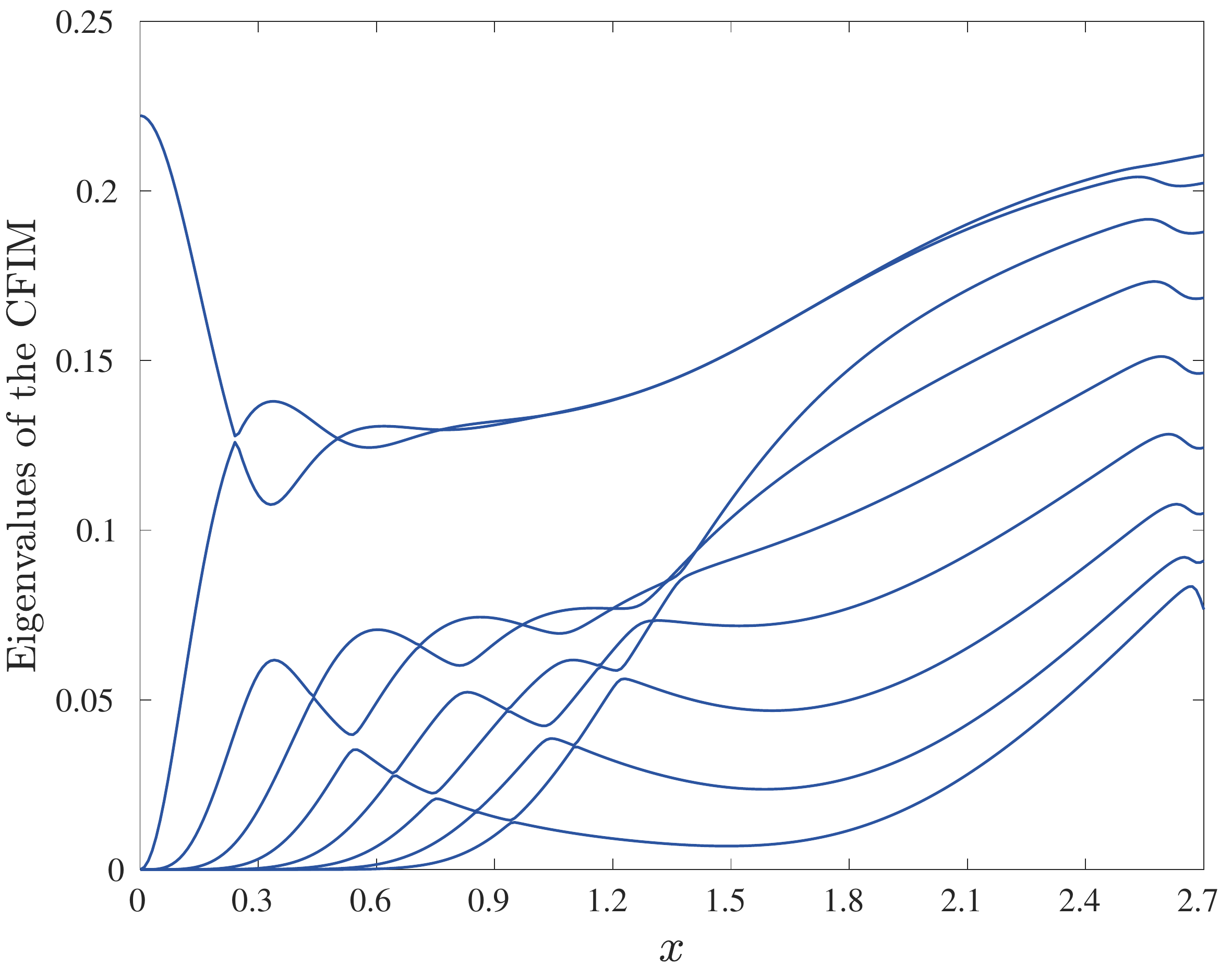}
 }
 \caption { The eigenvalues of the CFIMs in the case of 9 sources for SPADE (left) and direct imaging (right). The sources are positioned at \( \alpha_i = (i-5)x \).}
 \label{fig:CFIMs_eig}
\end{figure}

 \end{document}